\documentclass[%
 reprint,
 amsmath,amssymb,
 aps,
 prd
]{revtex4-2}

\usepackage{graphicx}
\usepackage{dcolumn}
\usepackage{bm}
\usepackage{color}
\usepackage{hyperref}
\hypersetup{
    colorlinks=true, 
    linktoc=all,     
    linkcolor=blue,  
}
\usepackage{caption}
\usepackage{subcaption}

\begin{document}

\preprint{APS/123-QED}

\title{Examining Temporal Variation of the Fermi Coupling Constant \\ using SNe Ia Light Curves}

\author{Akshay Rana}
\email{aksrana92@gmail.com}
\email{akshay@ststephens.edu}
\author{Vedanta Thapar}%
\author{Hariprasad S.V.} 
\author{Sandra Elsa Sanjai}

\affiliation{%
 St. Stephen’s College, University of Delhi, Delhi 110007, India
}%





\begin{abstract}
In standard model, the Fermi coupling constant, $G_F$, sets the strength of electroweak decay. We attempt an approach to constrain the temporal variation of the Fermi coupling constant $G_F$.  To probe it, Type Ia supernovae (SNe Ia) light curves are being used as a source of reliable primordial nucleosynthesis events across the redshifts. We utilized studies suggesting that in the initial phase after the SNe Ia explosion, the electroweak decay of $^{56}Ni \rightarrow ^{56}Co \rightarrow ^{56}Fe$ is the key contributor to  powering the SNe Ia light curve. We hence used the Pan-STARRS supernovae catalog having 1169 supernovae light curves in $g$, $r$, $i$, and $z$ spectral filters. The post-peak decrease in the apparent magnitude of light curves (in the rest frame of SNe) was related to the electroweak decay rate of primordial nucleosynthesis. Further, the decay rate relates to $G_F$. To keep the analysis independent of the cosmological model, we used the Hubble parameter measurement and a non-parametric statistical method, the Gaussian Process. Our study suggests a small yet finite temporal variation of $G_F$ and puts a strong upper bound on the present value of the fractional change in the Fermi coupling constant i.e; $\dfrac{\dot G_F}{G_F}\big\rvert_{z=0} \approx 10^{-11} yr^{-1}$ using datasets spread over a redshift range $0<z<0.75$.
\end{abstract}

\maketitle


\section{\label{intro} Introduction}

Even a glance at the different sectors of physics can give us a long list of quantities, known as \textit{fundamental constants}. Some of these are the universal gravitational constant ($G_N$),  speed of light $(c)$, reduced Planck's constant $(\hbar)$, the fine structure constant ($\alpha$), Fermi coupling constant ($G_F$), other gauge and Yukawa coupling constants. As our theoretical understanding of the laws of nature depends on these quantities, they are also known as constants of nature.  Traditionally, the fundamental constants were assumed to be constant at all locations and epochs of the space-time \cite{RevModPhys.93.025010,2017martines, 2019martines,Duff:2014mva,2002duff,2011uzan,2011chiba,frank}. In 1937, Dirac was the first one to question this assumption. In his pioneering work, \textit{Large Number hypothesis}, he speculated on the spatial and temporal variation of fundamental constants at cosmological scales \cite{1937Natur.139..323D,1938RSPSA.165..199D,kothari,chandrashekhar}. Space-time variation of these fundamental constants like: $c$, $G_N$, $\alpha$, $G_F$ etc, presents a possible window for the extension of theoretical frameworks beyond the standard cosmological and particle physics models \cite{Ellis_2009,Joyce_2015,Uzan_2003,2004PhT....57j..40O}. Hence, it is essential to study this variation of the observed values of these constants which can  highlight the underlying phenomena and may signal new exotic physics scenarios.\\

One of the major challenges in studying space-time variation of fundamental constants is that even if any such variation is present, it would be nearly impossible to detect it through Earth-based laboratory experiments. The reason is, the changes involved in such a variation would be extremely small and hence any laboratory experiment would require non-viable levels of accuracy to detect them. However, astrophysical objects and events spread  over length scales as vast as thousands of mega-parsec and time scales as large as Giga-years emerge as a nature-based testing lab for studying the constancy of fundamental constants. In recent times, unprecedented developments in the field of observational and experimental capabilities have opened new avenues to test  the  temporal variation in the current values of these constants using astronomical observables. One may locate many such attempts in literature for different fundamental constants with different observables \cite{Martins_2016,Salzano_2017,Wang_2020,PhysRevD.41.1034,1986PhRvD..33..869K}. \\

In this work, we probe any hint of the temporal variation of the Fermi coupling constant, $G_F$, using primordial nucleosynthesis observations of Type Ia supernovae (SNe Ia). The Fermi coupling constant is first introduced in the four-particle coupling explanation of the beta decay in Fermi's weak interaction theory. The same phenomenon was later analysed in the strong electroweak theory and this interaction is found to be governed by the exchange and production of $W$-bosons. The Fermi coupling constant determines the strength of the weak interaction and the electroweak rate of decay and in Planck's unit it is defined as $G_F= \dfrac{\sqrt{2}}{8} \dfrac{g^2}{M_W^2}$. Here $g$ stands for the standard model's SU(2) gauge coupling constant and $M_W$ represents the mass of the $W$-bosons. Unlike gluons and photons which are massless, intermediate $W$-bosons are extremely massive. In the standard model, especially the Higgs mechanism, the mass of the bosons is given by $M_W= \dfrac{1}{2}\nu g$ where $\nu$ represents the vacuum expectation value ($vev$) of the Higgs boson. On replacing the $M_W$ value in $G_F$ expression, we find that the Fermi coupling constant is independent of $g$, instead, it is a direct measure of $vev$.\cite{GRUNEWALD1999125,alt,chris}. \\

The Higgs mechanism is responsible for the mass generation of matter excitation that includes bosons, fermions, and quarks. The masses of the quarks and the fermions depend on $vev$, as well as on the Yukawa coupling corresponding to quarks and fermions. Hence any spatial or temporal variation in the $G_F$ value will change the mass of fermions and quarks. In turn, it will lead to several inevitable changes to our current understanding of the standard model of particle physics. Some prominent phenomena will be as follows: a) the variation of fermion mass will indicate the change in electron's mass $m_e$, b) the quark mass change will affect the neutron-proton mass difference and, c) the variation in the mass of the pion  will eventually impact the strong nuclear forces \cite{ds1988,1986PhRvD..33..869K}. Although no credible evidence of mass variation of these fundamental particles has been observed, there is a possibility that along with the $vev$, Yukawa couplings also change to keep the mass of the corresponding particle constant.  Besides this, there are theories in literature that suggest Lorentz violation as an alternative to the Higgs mechanism\cite{lor1, lor2}. All these possibilities could open a window to new physics and hence these factors make it crucial to probe the constancy of $G_F$ at cosmological scales.\\

One of the most promising ways of determining the present value of the Fermi coupling constant, $G_F$, is through the precise measurement of the mean lifetime of muons. The presently accepted value of the Fermi coupling constant is measured through a laboratory setup of a scintillator detector array and the value is found to be $G_F= 1.1663787(6) \times 10^{-5}$ GeV$^{-2}$\cite{bound1, bound2, bound3}. Though this method provides a very precise estimate of the current value of $G_F$, it doesn't indicate anything about the spatial or temporal variation in the value of the constant. Hence, for that purpose, a cosmological nucleosynthesis-based probe  is needed.  One such celestial observatory is found in the form of Type Ia supernovae (SNe Ia) which act as a laboratory of primordial nucleosynthesis and hence play a crucial role in the study of electroweak interactions across the universe \cite{2009liebend}. \\

 In literature, Kolb, Perry \& Walker (herein KPW1986) were one of the first to study the impact of changing fundamental constants on the standard model of primordial nucleosynthesis.   They established that an increasing $G_F$ will reflect as an increased rate of weak interaction in nucleosynthesis \cite{1986PhRvD..33..869K}. Following KPW1986, Dixit \& Sher (1988) studied the variation of $G_F$ and  its impact on primordial nucleosynthesis\cite{ds1988}.
   Further, Scherrer \& Spergel (1993) find that if we consider the $G_F$ variation to be independent of any change in the fermion masses, then the best constraint on temporal and spatial variation in $G_F$ would come from the SNe Ia light curves\cite{ss1993}. With a single observation of SN1988U at a redshift $z=0.3$ and length scale of 30 Mpc, \cite{sn1988}, Scherrer \& Spergel (1993) hinted that $G_F$ could vary by at most $5\%$ at such cosmological scales. Later, Ferrrero \& Altschul (2010) used the light curves of SNe Ia from a CfA supernova archive in the redshift range $0.0039<z<0.0240$ and find a fractional change in the Fermi coupling constant of the order of $10^{-9}$ yr$^{-1}$ as a $2\sigma$ bound \cite{ferrero}. Karpikov et al. (2015)  used the publicly available light curves of JLA compilation of SNe Ia and probe the limits on the nuclear decay rates of primordial nucleosynthesis at redshifts as large as $z \approx 1$ \cite{karpikov15}.\\
   
  In this work, we are studying the temporal  variation of the Fermi coupling constant by determining the weak decay rates of primordial nucleosynthesis processes in the universe. For that, we proposed a new approach where we used the publicly accessible 1169  photometrically classified Pan-STARRS supernovae compilation. Approximately 95\% of these supernovae are SNe Ia with the remaining being Core-Collapse(CC) ones \cite{2018ApJ...857...51J}. We used the PSNID classifier to analyse all of the supernovae' light curves in order to ensure that those selected for analysis are of Type Ia.
   In Section \ref{section2} of the manuscript, we elaborate on the methodology used, Section \ref{section3} discusses the dataset, selection criterion, and the various steps of the analysis. Section \ref{section4} highlights the results while a detailed discussion of the outcomes of the analysis can be found in Section \ref{section5}.
   
\section{\label{section2} Methodology}

As pointed out in KPW1986, an independent estimate of the measurements of the rate of weak interactions at different redshifts can be used to probe the temporal variation of $G_F$ \cite{1986PhRvD..33..869K}. Hence, we study the decay rate of weak interaction in primordial nucleosynthesis-based astrophysical processes taking place at different cosmic epochs. SNe Ia are one such widely distributed high-redshift  astrophysical objects known for their universality in the explosion process, evolution, and dynamics \cite{sup0,sup1}. The widely accepted theory of the SNe Ia explosion is based on the luminous stellar thermonuclear explosion that occurs when the mass of a white dwarf reaches up to the Chandrashekhar mass limit by accreting mass from an external binary source. This causes the white dwarf's gravity to dominate over electron degeneracy pressure forcing it to collapse into itself. This compression causes heating of the core and uncontrolled nuclear reactions to occur. The energy released in turn causes expansion, reducing the density allowing burning which quickly leads to detonation. The high temperatures and pressures set the stage for quick thermonuclear fusion resulting in most of the material (Carbon, Oxygen, etc.) being burnt to Ni and Si. This theory is supported by observations of spectra of SNe Ia which are dominated by these elements among others\cite{Hoflich:2003bg, ch1,ch2}.\\

The most common observation of the supernovae is its light curve: a plot of the flux or apparent magnitude of the supernova vs time (in days) which contains information about the energy released from the supernova as it occurs. Observations and simulations suggest that the hydrodynamic and nuclear burning processes last only approximately for few minutes after the SNe Ia explosion. The subsequent luminosity is powered entirely by the decay of radioactive elements synthesized in the explosion, in particular, the doubly magic nuclei of $^{56}Ni$ which further decays through a chain reaction as $^{56}Ni \to ^{56}Co \to ^{56}Fe $ \cite{ch1}. It turns out that energy radiated from supernovae can be measured for several years after the explosion. However, it peaks roughly after 18-20 days of explosion and then starts decaying. In the initial stage of the explosion, the radioactive decay of $^{56}Ni \to ^{56}Co$ is the primary source of the energy radiated by the SNe Ia \cite{ni,Diehl_2014}.  However $^{56}Co$,  still being radioactive, decays to stable $^{56}Fe$ over a far longer time and is solely responsible for the late time energy of the SNe Ia light curve. \\

\noindent Radioactive $^{56}Ni$ transforms via `Electron Capture' (EC) into radioactive $^{56}Co$. 
$$^{56}Ni \Longrightarrow ^{56}Co + \gamma + \nu_e$$

   Further, radioactive  $^{56}Co$, the product of $^{56}Ni$ decay, transforms into a stable isotope $^{56}Fe$ either through the process of `Electron Capture' (EC) or `Positron Decay' (PD).

$$
^{56}Co \Longrightarrow
\begin{cases}
^{56}Fe + \gamma + \nu_e & \text{EC}\\
^{56}Fe + e^+ + \gamma + \nu_e & \text{PD}
\end{cases}
$$

The  power radiated by SNe Ia is precisely known to be an exponentially decaying function of the rest frame time ($\tau$) given by \citep{decay_exp,Sukhbold}
  
\begin{equation}
L_{\gamma}= \dfrac{M_{Ni}}{M_{\odot}} (C_{Ni} \exp^{-\tau/\tau_{Ni}}+ C_{Co} \exp^{-\tau/\tau_{Co}}) \textnormal{  ergs s}^{-1}
\end{equation}
  
where  $M_{Ni}$ and $M_{\odot}$ are mass of $^{56}Ni$ and solar mass respectively.  $C_{Ni} \approx 6.45 \times 10^{43}, C_{Co} \approx 1.45 \times 10^{43}, \tau_{Ni}= 8.77 \textnormal{ days}$ and $\tau_{Co}= 112.1 \textnormal{ days}$ \cite{thalf,ni,co_decay,decay_exp} . Hence for SNe Ia, it would be safe to assume that after reaching the peak of the light curve even for roughly the next 60 days, the energy radiated and hence the shape of the light curve is primarily governed by the electroweak radioactive decay of $^{56}Ni $ and $ ^{56}Co$ \cite{Hoflich:2003bg, lc, lc2, ni}. These nuclear decay processes are governed by the weak interaction. The $\gamma$ photons arising from the decay interact with the ejecta by photo-electric absorption, Compton scattering, and pair production which is further thermalized to emit UV, optical, and IR photons.\\
  
  The near equality of the masses of the white dwarf progenitors gives all SNe Ia similar energy outputs and makes them extremely important standard candles in cosmology. As all SNe Ia light curves decay uniformly, any variation in the decay rate of the different supernovae at different redshifts and locations will be a measure of the variation of the weak interaction decay rate of primordial nucleosynthesis.  As the decay rate ($\Gamma$) in weak interaction is directly related to the Fermi coupling constant ($G_F$) as $\Gamma \propto G_F^2$, it will give us a handle to probe the constancy of $G_F$ \cite{decayofni}. Hence the fractional change in $G_F$ would be given by 
  
\begin{equation}
    \frac{\dot G_F}{G_F} =\dfrac{1}{2} \left( \frac{\dot \Gamma}{\Gamma}\right) 
    \label{maineq}
\end{equation}

  As the radioactive decay of $^{56}Ni$ and  $^{56}Co$ drop exponentially, the flux will also drop exponentially. Further, the flux and apparent magnitude, $m$, of SNe Ia are related by a logarithmic relation \cite{ferrero,karpikov15}. Hence the decay rate of the weak interaction process  will be proportional to the slope or rate of change of apparent magnitude ($dm/d\tau$)  of the SNe Ia light curve during the rest frame time $\tau $ $\approx$ 60 days from the peak i.e. 
 
\begin{equation}
    \Gamma \propto \dfrac{dm}{d\tau}
    \label{gammapropm}
\end{equation}

Further, we can find the relative change in the decay rate of electroweak interaction over the different epochs of the universe by analysing the light curves of  SNe Ia at different redshifts.  It is given by $\dfrac{\dot \Gamma}{\Gamma}$ using Eq. \ref{gammapropm} as follows:

\begin{equation}
    \frac{\dot \Gamma}{\Gamma} = \dfrac{1}{dm/d\tau}  \dfrac{d(dm/d\tau)}{dt} 
 \label{eq_2}
\end{equation}

 Here it is important to point out that the time  $\tau$ is the rest frame time during the supernovae explosion while the time $t$ is a measure of cosmological time scales. Here onward, for simplification, we define $dm/d\tau= m_{\tau}$. Hence the relation will modify as;
 
 \begin{equation}
    \frac{\dot \Gamma}{\Gamma} = \frac{1}{m_{\tau}} \frac{d m_{\tau}}{dt} = \frac{1}{m_{\tau}} \frac{d m_{\tau}}{dz} \frac{dz}{dt}
 \label{eq_2i}
\end{equation}

Here $z$  is the redshift and $t$ is the cosmological time scale. The $\dfrac{dt}{dz}$ parameter can be related to Hubble parameter $(H(z))$ by the standard relation for an expanding universe, given by $\dfrac{dt}{dz}= \dfrac{-1}{(1+z)H(z)}$ and hence the modified expression for Eq. \ref{eq_2i} will be

\begin{equation}
  \frac{\dot \Gamma}{\Gamma} = -\frac{(1+z)H(z)}{m_{\tau}} \frac{d m_{\tau}}{dz} 
  \label{dm/dz}
\end{equation}

Hence to measure the fractional change in decay rate at different epochs, we need to find the independent measurements of  $m_{\tau}$, $\dfrac{d m_{\tau}}{dz}$, and $H(z)$ at different redshifts. In order to obtain the first two quantities, we used SNe Ia light curve, and for $H(z)$ measurements, we used the data based on cosmic chronometers. The details of the datasets are given in the next section.

\section{\label{section3}Dataset and Analysis}

In this work, we used the Pan-STARRS supernovae catalog compiled by Jones et al.(2018), having 1169 supernovae in the redshift range $0.02<z<0.95$ \cite{2018ApJ...857...51J}\footnote{For data follow the link: \href{https://archive.stsci.edu/prepds/ps1cosmo/jones_datatable.html}{https://archive.stsci.edu/prepds/ps1cosmo/jones\_datatable.html}}. All the supernovae in this catalog are not spectroscopically classified, and photometric classification suggests that approximately 95\% of these  are SNe Ia, with the remainder being Core-Collapse (CC) supernovae.  This catalog contains the detailed early time light-curve observation for all the supernovae in four spectral filters ($g,\ r,\ i,\ z$ filters) based on the SDSS filters system. These filters ($griz$ herein) correspond to a central frequency 4686$\text{\AA}$, 6166$\text{\AA}$, 7480$\text{\AA}$, and 8932$\text{\AA}$ respectively\cite{filter}. We preferred to work with Pan-STARRS supernovae catalog over the Pantheon compilation of SNe Ia as in the Pan-STARRS catalog, all the supernovae had been observed through the same survey having a uniform systematic error. While Pantheon compilation of SNe Ia has all spectroscopically classified SNe Ia observations but from nine different surveys which will introduce a different level of systematic errors. Besides it, in the Pantheon compilation, all SNe Ia are not observed across the same filters needed for this analysis \cite{scolnic2018}.\\

PSNID\cite{psnid} matches the observed light-curve  to simulated Type Ia and core-collapse (CC) light curves  generated by the  SALT2 Model and  over 51 CC SN templates. This returns a probability that a given SN is of Type Ia based on the goodness of  fit and $\chi^2$ of the SALT2 fit performed by SNANA. We implemented a selection criterion and  classify only SNe Ia with $PSNID \geq 0.95$ as Type Ia to be used in our analysis. This leaves us with approximately 70\% of the dataset (814 SNe Ia) with significant confidence that all these light curves correspond to SNe Ia. 

\subsection{\label{dotmz}Estimating the $m_{\tau}$ vs $z$ }

\begin{figure*}[t]
\centering
\includegraphics[width=0.8\textwidth]{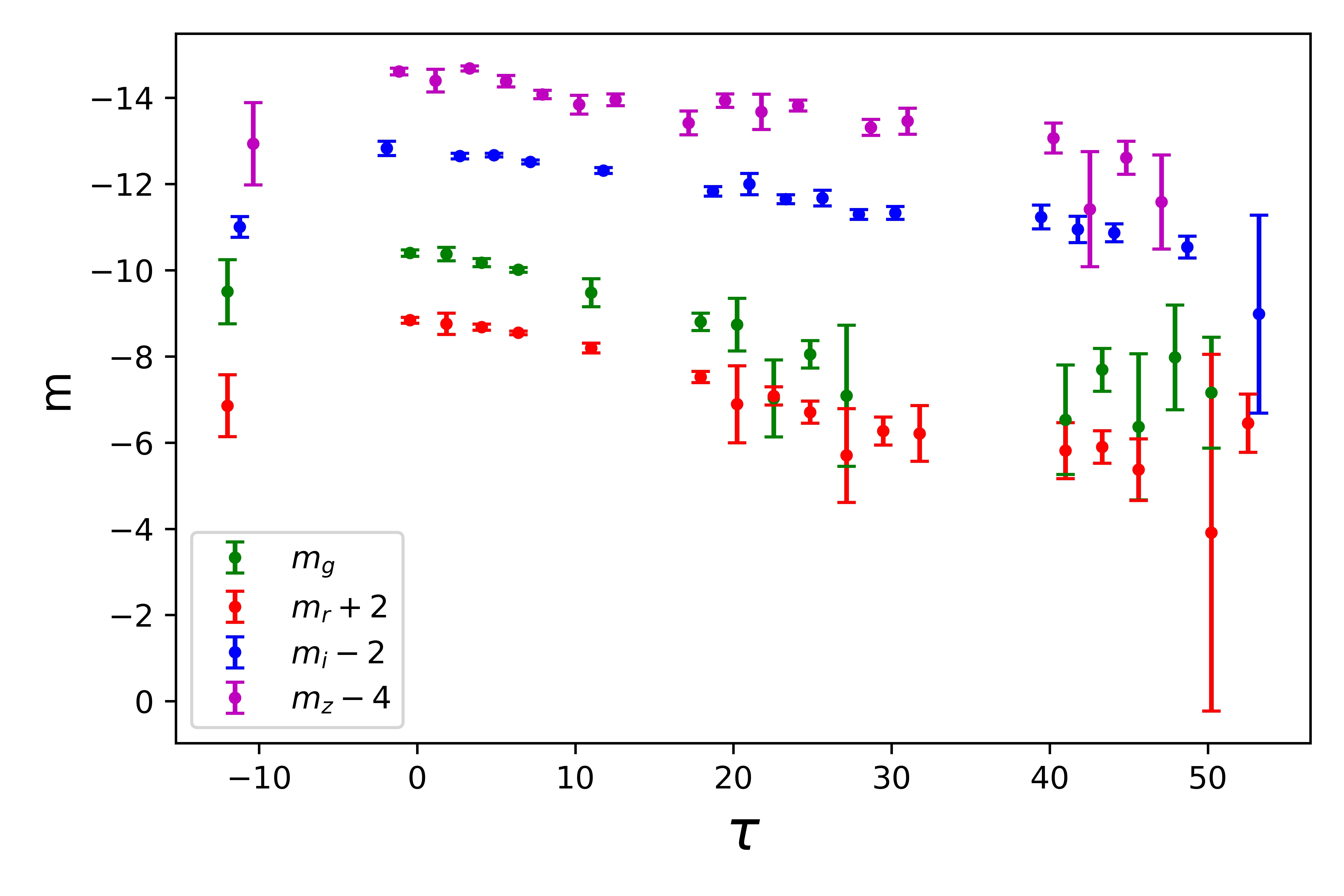}
\caption{ Light Curve of  SNe Ia \texttt{psc170428} around the peak in all four ($griz$) filters. The parameter $\tau$ represents the rest frame time and $\tau=0$ indicates the peak of the light curve. On vertical axis, the apparent magnitude $(m)$ for the filters has been re-scaled to avoid the overlap of data points of different filters. }
\label{fig:my_label}
\end{figure*}

As discussed in Section \ref{section2}, now our aim is first to isolate the light-curve observations of each supernova from the peak to $\approx$ 60 days in the rest frame of the SNe Ia in question. To explain all the intermediate steps of analysis, we chose a supernova having ID \texttt{psc170428}.  Our first step was to divide our data points into four different datasets corresponding to each filter (\textit{griz} filters). To do this we first removed all the data points for which the magnitude was undefined and points for which the error is greater than magnitude order $\pm 5$. Then depending on the values of the \texttt{FLT} key, we divided the dataset into four lists. Further, the light curve of each SNe Ia is transformed from the observer's frame to rest frame of the source SNe Ia. This is obtained by dividing the observer's frame time by a factor of $(1+z)$. The evolution of the example SNe  in corresponding rest frame time($\tau$) and peak scaled as reference point is shown in Fig. \ref{fig:my_label}. 

The peak MJD of the SNe Ia is taken to be that provided in the data and rescaled as above to rest frame. This value is found by fitting a light curve  using the SALT2 algorithm provided by the SNANA package. We further confirmed these values by independently calculating the peak using the SNCOSMO package which also uses the SALT2. Both packages were found to produce similar results and the SNANA values were used to maintain consistency. A benefit to using SALT2 is that it applies corrections to the spectral sequence as a function of phase, wavelength and a stretch factor in such a manner that the spectra integrated over the response functions match the observed light-curves. As the K-corrections are naturally built into the  SALT-2 model hence it provides us with a tool to fit the observed light-curves without correcting the data points (see e.g. core Nugent et al. (2002) \cite{NO2} for the definition of K-corrections) \cite{guy2007}.The SALT2 algorithm is widely used and has been quite successful when applied to estimating SNe Ia distances at high redshift.

\begin{figure*}[]
     \centering
     \subfloat[$g$ filter Fit]{\includegraphics[width=8cm]{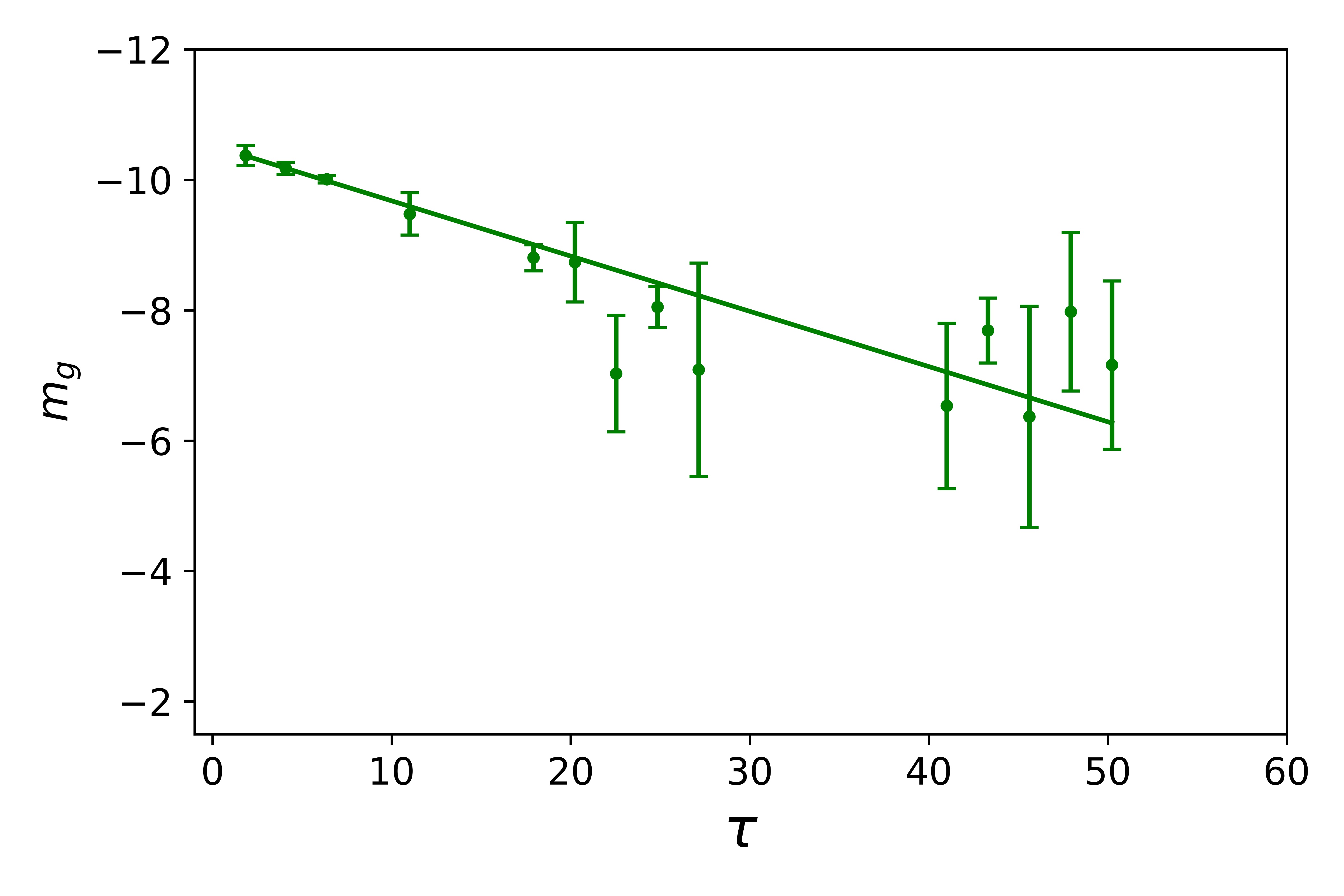}}
     \subfloat[$r$ filter Fit]{\includegraphics[width=8cm]{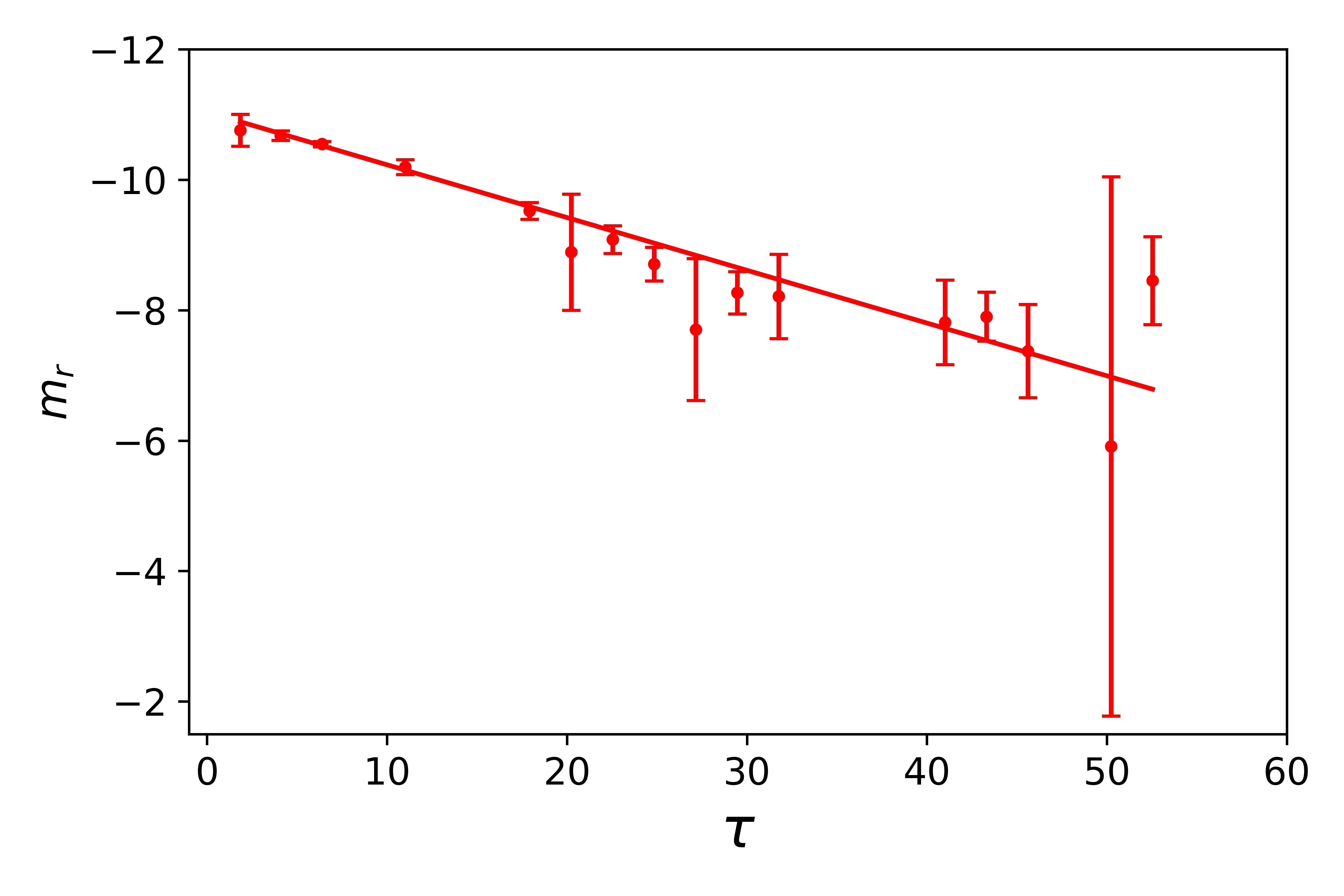}} \\
     \subfloat[$i$ filter Fit]{\includegraphics[width=8cm]{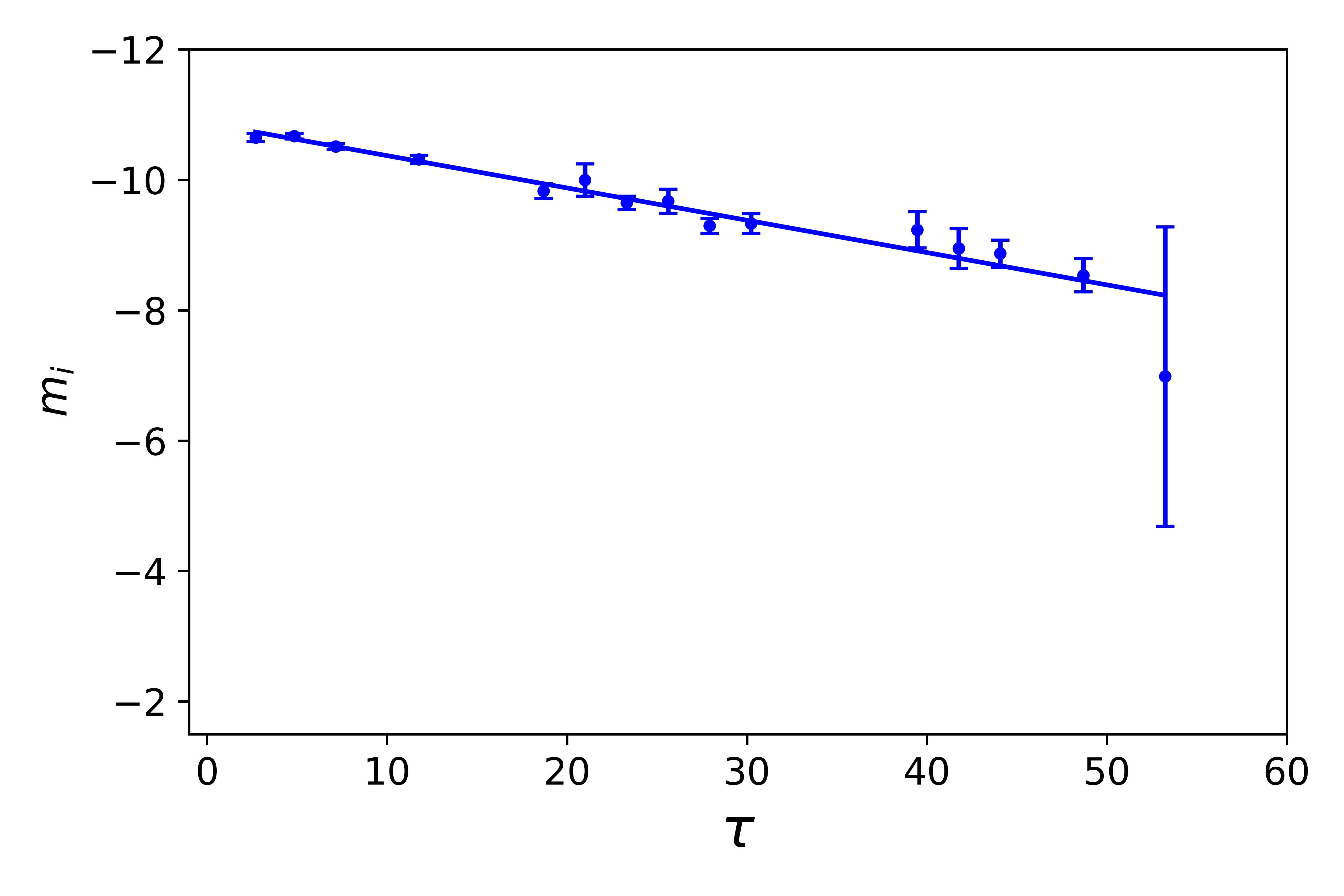}}
     \subfloat[$z$ filter Fit]{\includegraphics[width=8cm]{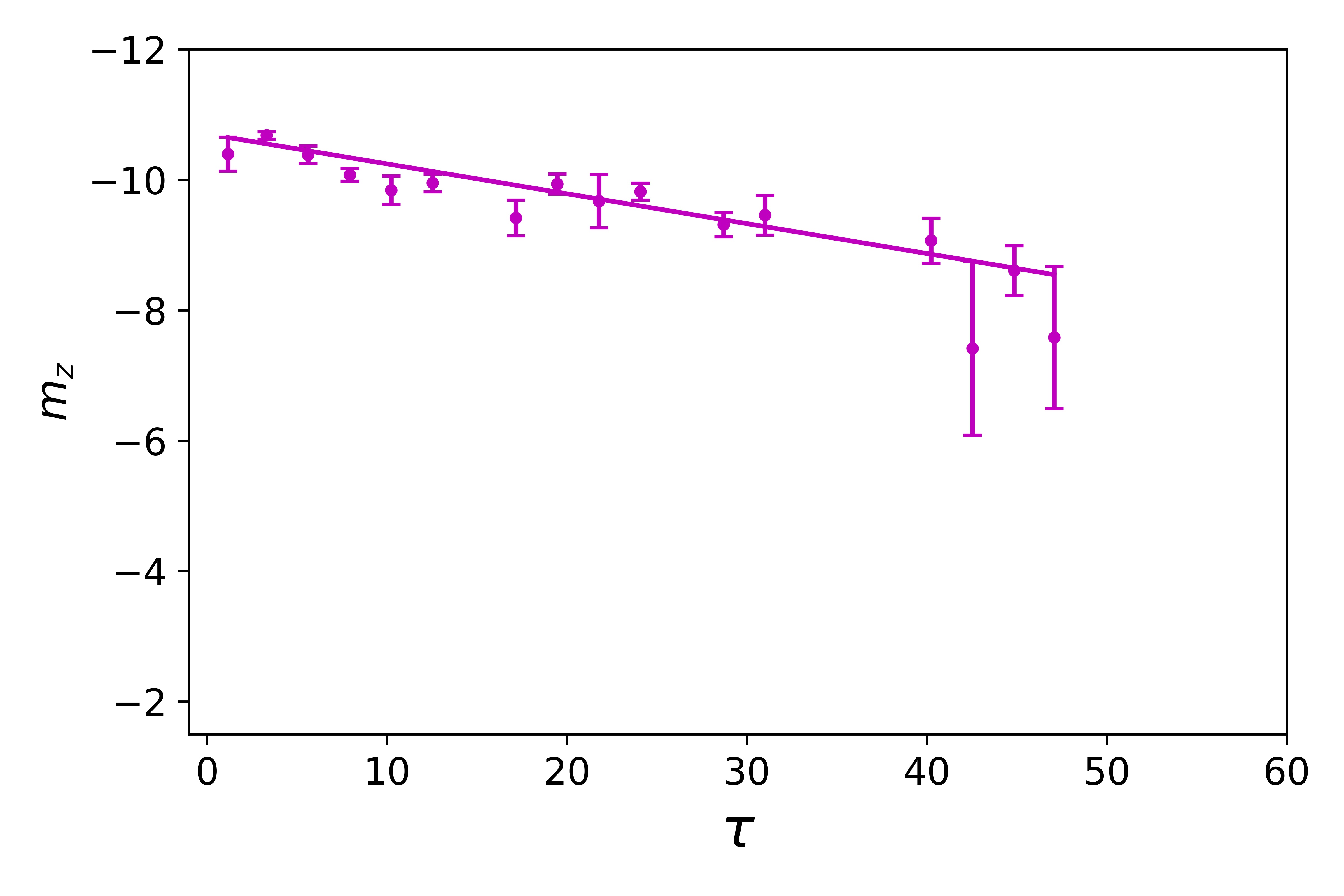}}
     \caption{This plot contains the observed measure of apparent magnitude of SNe Ia \texttt{psc170428} light curve for $griz$ filters during a time period of 60 days starting from the peak of the light curve. The straight solid line indicates the linear best fit to the data. }
     \label{fig:grizcurve}
 \end{figure*}
 
 We now consider only the points for $\tau \in [0, 60]$ and aim to estimate the rate of decay of the light curve. To measure $m_{\tau}$ (ie. $dm/d\tau$) within the region of interest we chose to fit a straight line to the data points. The reason for this is essentially visual, and further, on trying to fit alternative functions like higher order polynomial or non-polynomial functions like $f(\tau) = a + b\ln \tau$ or $f(\tau) = a + b\frac{\tau}{1+\tau}$, we found those to be inconsistent with the data. Hence we chose to fit straight line  $f(\tau) = a + b\tau$ to the selected $n$ datapoints of light-curve in the noted time period. In this way, the slope parameter $b$ is equivalent to $m_\tau$. The goodness of fit was measured using the $\Delta \chi^2$ method and the fitting was done by $\chi^2$ minimization.  The parameter $\chi^2$ is defined as follows:
 \begin{equation}
     \chi^2 = \sum_{i=1}^n \left(\frac{m_i - f(\tau_i)}{\sigma_i}\right)^2
 \end{equation}

\begin{figure*}[t]
\centering
\begin{subfigure}{.5\textwidth}
  \centering
  \includegraphics[width=1.0\linewidth]{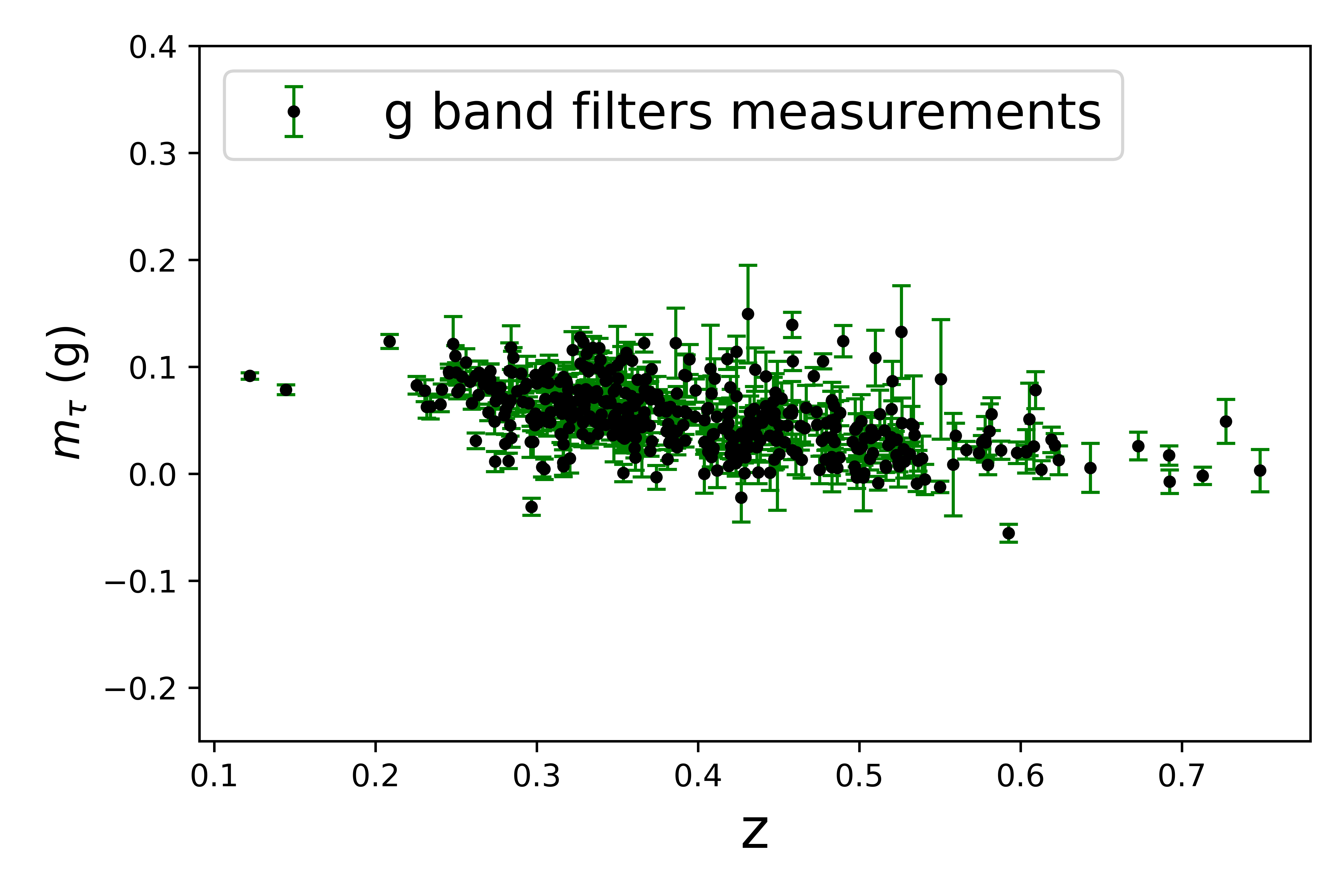}
  \caption{ $m_\tau$ vs $z$ measurement for g filter}
  \label{fig:sub15}
\end{subfigure}%
\begin{subfigure}{.5\textwidth}
  \centering
  \includegraphics[width=1.0\linewidth]{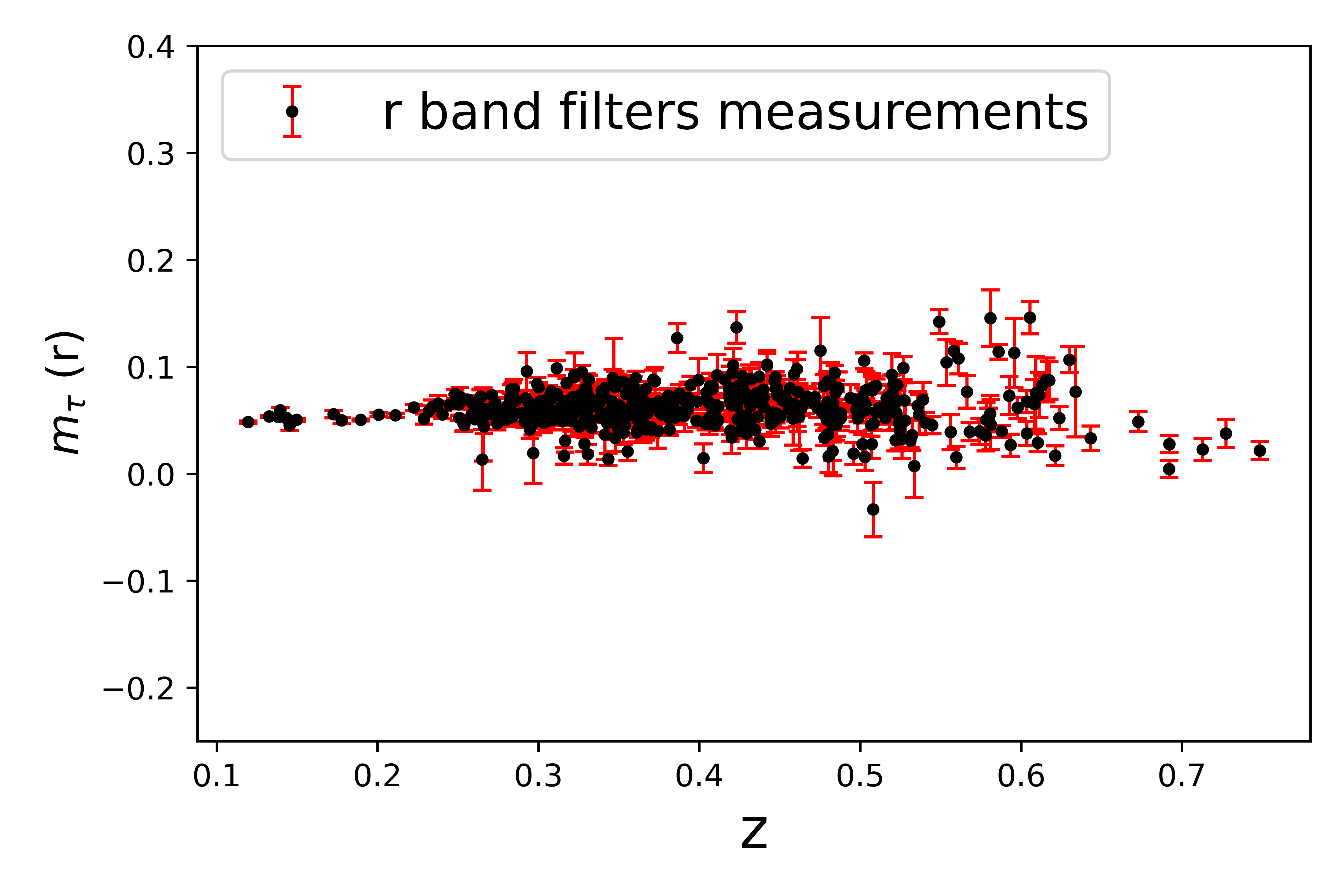}
  \caption{$m_\tau$ vs $z$ measurement for r filter}
  \label{fig:sub28}
\end{subfigure}

\begin{subfigure}{.5\textwidth}
  \centering
  \includegraphics[width=1.0\linewidth]{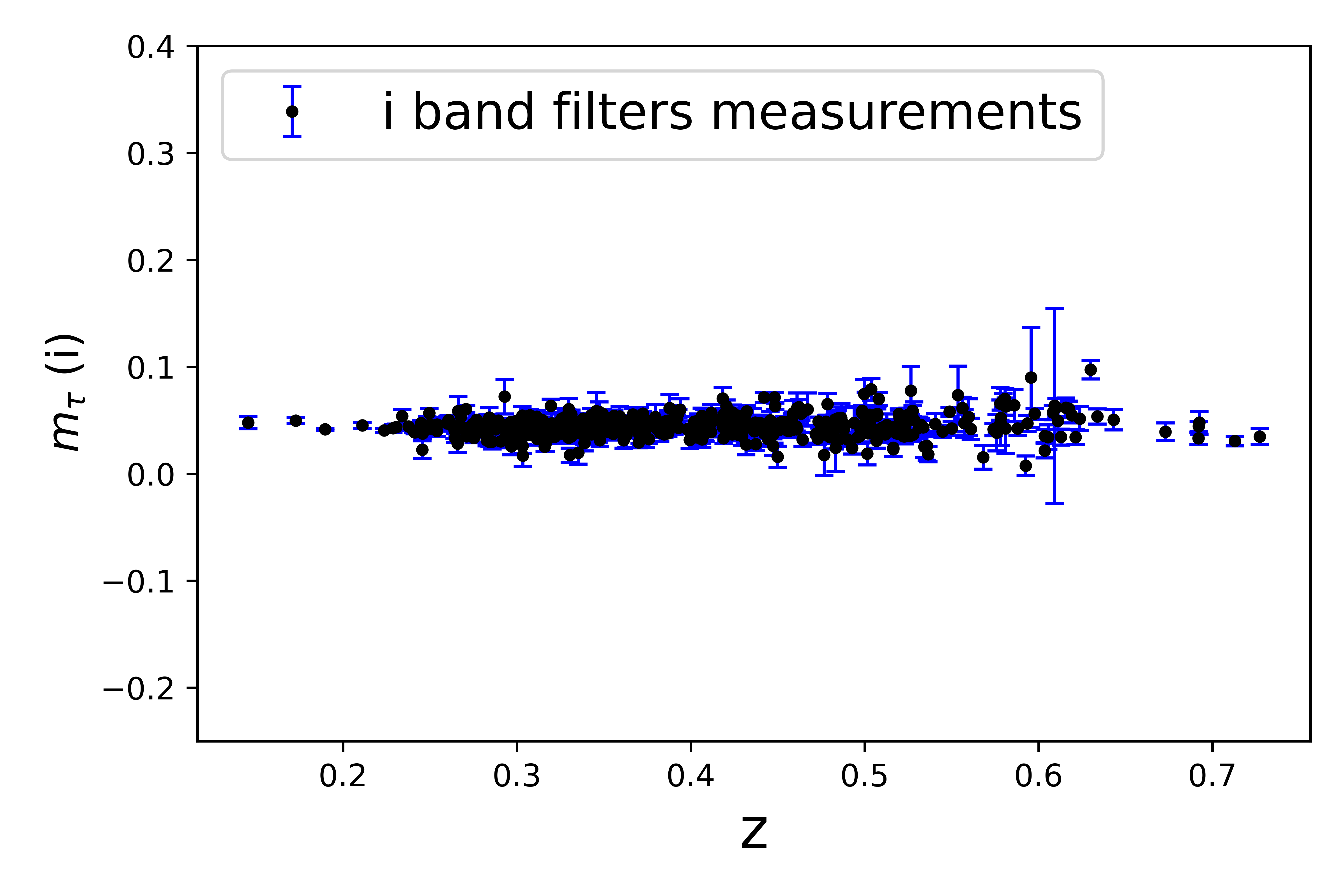}
  \caption{$m_\tau$ vs $z$ measurement for i filter}
  \label{fig:sub11}
\end{subfigure}%
\begin{subfigure}{.5\textwidth}
  \centering
  \includegraphics[width=1.0\linewidth]{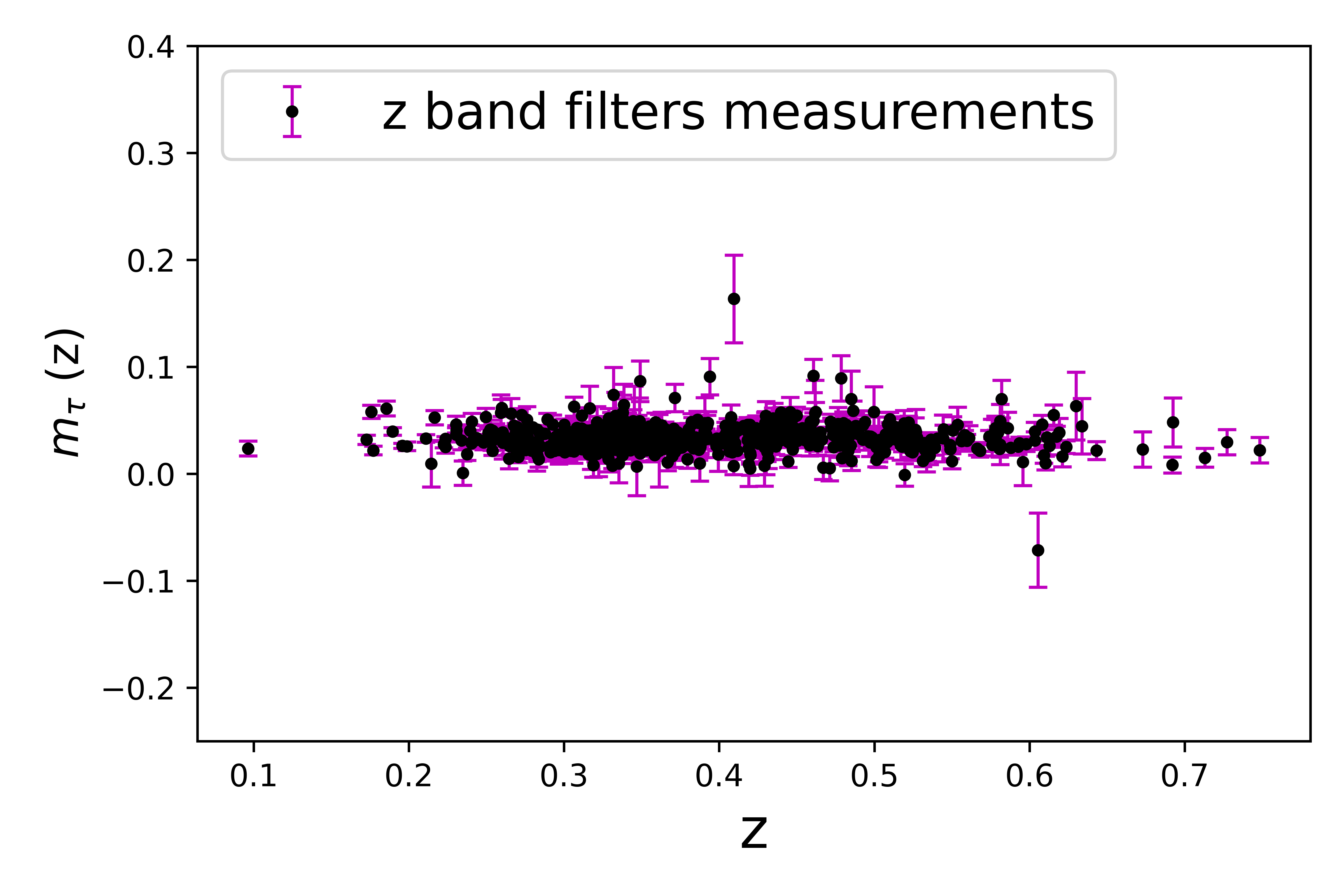}
  \caption{$m_\tau$ vs $z$ measurement for z filter}
  \label{fig:sub21}
\end{subfigure}

\caption{ Each subplot here contains the measure of $m_\tau  (=\frac{dm}{d\tau})$ vs $z$ for the specified filter($griz$ respectively) coming from the SNe Ia light curves spreaded over the redshift range $0<z<0.75$. }
\label{mdot}
\end{figure*}

In Fig. \ref{fig:grizcurve}, the straight line fit is shown only for the sample SNe Ia \texttt{psc170428}. In the similar fashion, we repeated the calculation for all selected 814 supernovae in the catalogue. To ensure the robustness of our analysis, we applied the following selection criterion on the selected data points for each SNe Ia. 
\begin{itemize}
    \item Minimum number of data points in the fitting region must be greater than or equal to 5.
    \item To ensure goodness of fit, the value of $\Delta \chi^2$  should be less than 2 i.e; $\Delta \chi^2 \leq 2$.
\end{itemize}
Finally, we are left with 429 measurements of $m_\tau$ for $g$-filter, 482 for $r$-filter, 535 for $i$-filter, and 622 for $z$-filter. We recorded the value of ${m_\tau}$ and associated error $\sigma_{m_\tau}$ for all the selected supernovae in the redshift range $0<z<0.75$ in all four filters (see Fig. \ref{mdot}).

\subsection{\label{subsec2}Obtaining $\dfrac{d m_\tau}{dz}$ using the Gaussian Process}

Once we find the $m_\tau$ vs $z$ measurements for the supernovae in different filters then the next step of the analysis is to obtain the $\frac{d m_\tau}{dz}$ and the value of the Hubble parameter $H(z)$ at all the redshifts of the supernovae. We did the following task using a model independent non-parametric smoothing and reconstructing method, namely, the Gaussian Process (GP) \cite{rasmussen_gp,seikel_gp,gp2}. It is a Bayesian approach based on the multivariate normal distribution. In a Gaussian Process, given some data points, we want to fit a function to the data. However, instead of using some complicated parametric relationship, we estimate a distribution of functions over whole data characterised by a mean function and a covariance matrix. This method comes with a few underlying assumptions as all the data points are independent and belong to the same population however neighboring points (let us assume two redshifts $z$ \& $z'$) do correlate due to their nearness. This correlation is taken care of through a kernel function given by
\begin{equation}
    k(z,z')= \sigma_f^2 \exp - \dfrac{(|z-z'|)^2}{2l^2}
\end{equation}

\begin{figure*}[t]
\centering
\begin{subfigure}{.5\textwidth}
  \centering
  \includegraphics[width=1.0\linewidth]{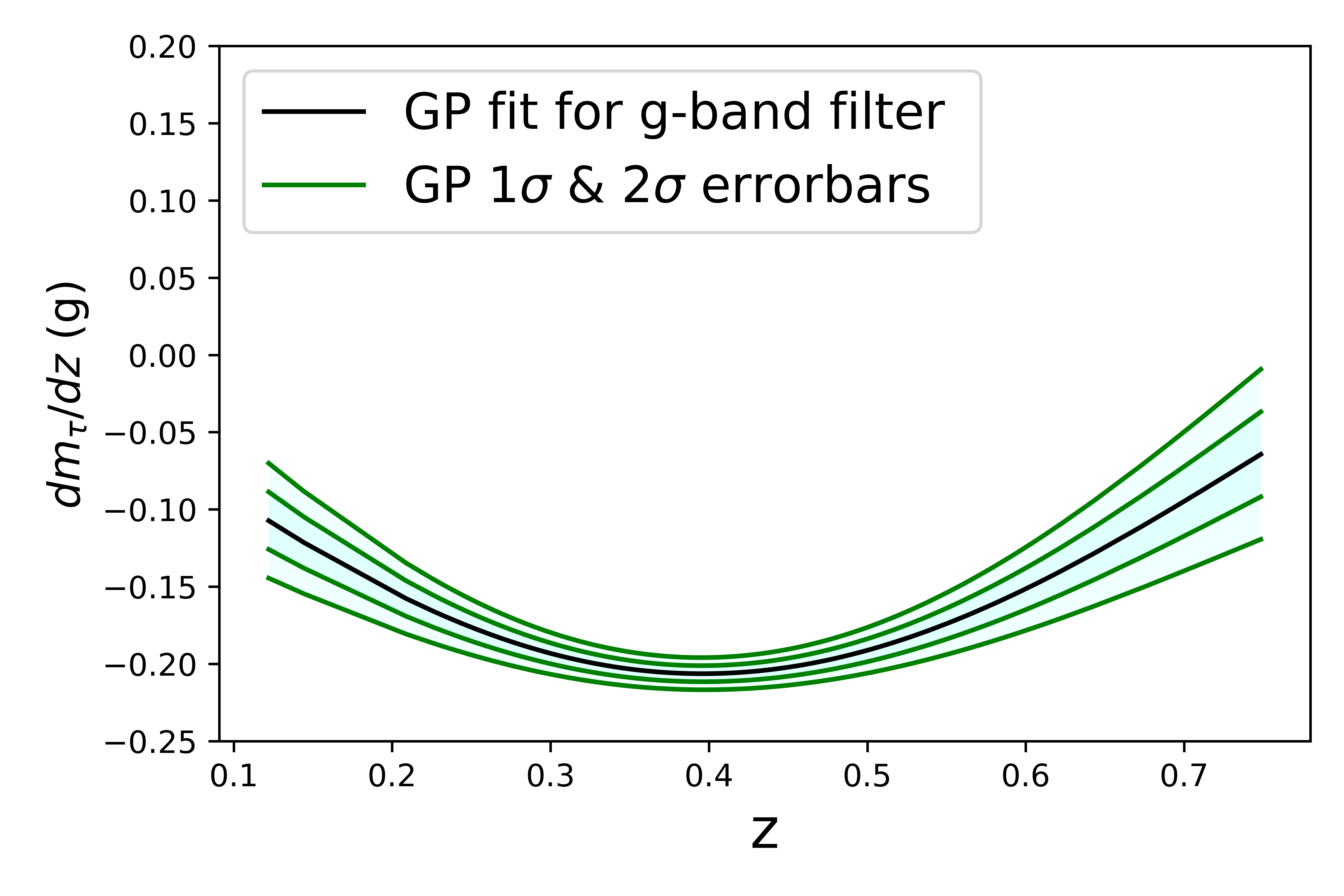}
  \caption{ $\frac{d m_\tau}{dz}$ vs $z$ measurement for g filter}
  \label{fig:sub1x}
\end{subfigure}%
\begin{subfigure}{.5\textwidth}
  \centering
  \includegraphics[width=1.0\linewidth]{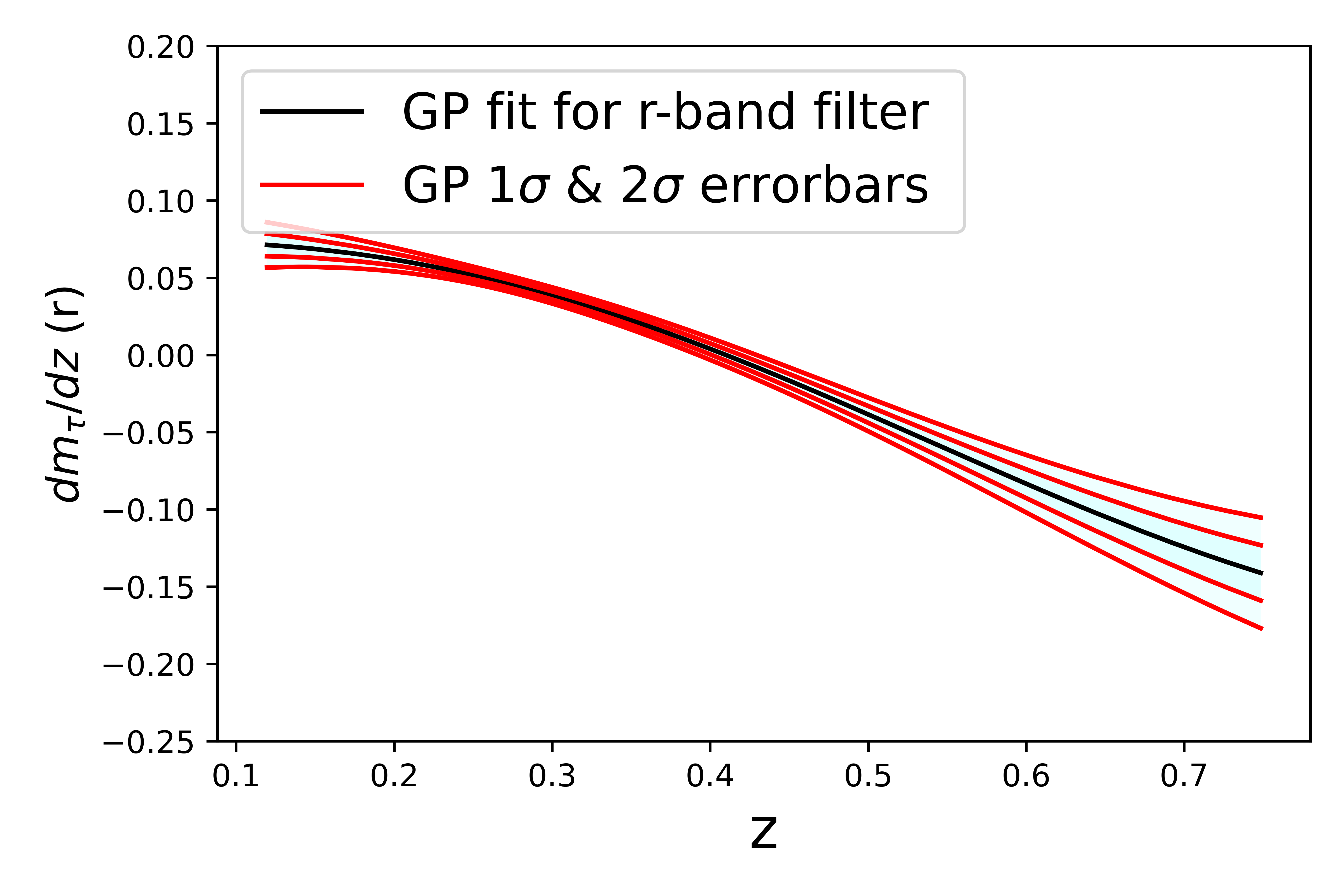}
  \caption{$\frac{d m_\tau}{dz}$ vs $z$ measurement for r filter}
  \label{fig:sub22}
\end{subfigure}

\begin{subfigure}{.5\textwidth}
  \centering
  \includegraphics[width=1.0\linewidth]{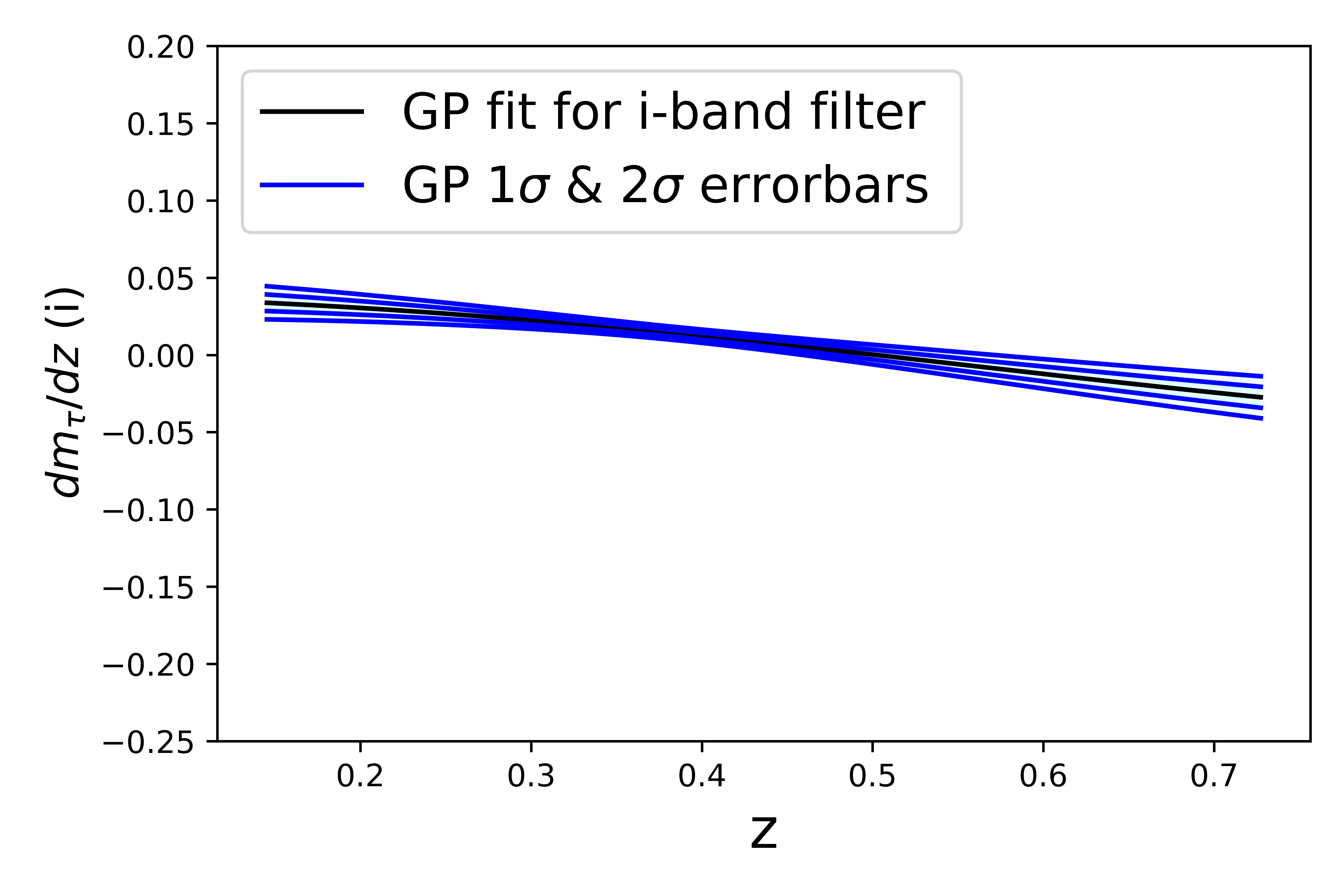}
  \caption{$\frac{d m_\tau}{dz}$ vs $z$ measurement for i filter}
  \label{fig:sub12}
\end{subfigure}%
\begin{subfigure}{.5\textwidth}
  \centering
  \includegraphics[width=1.0\linewidth]{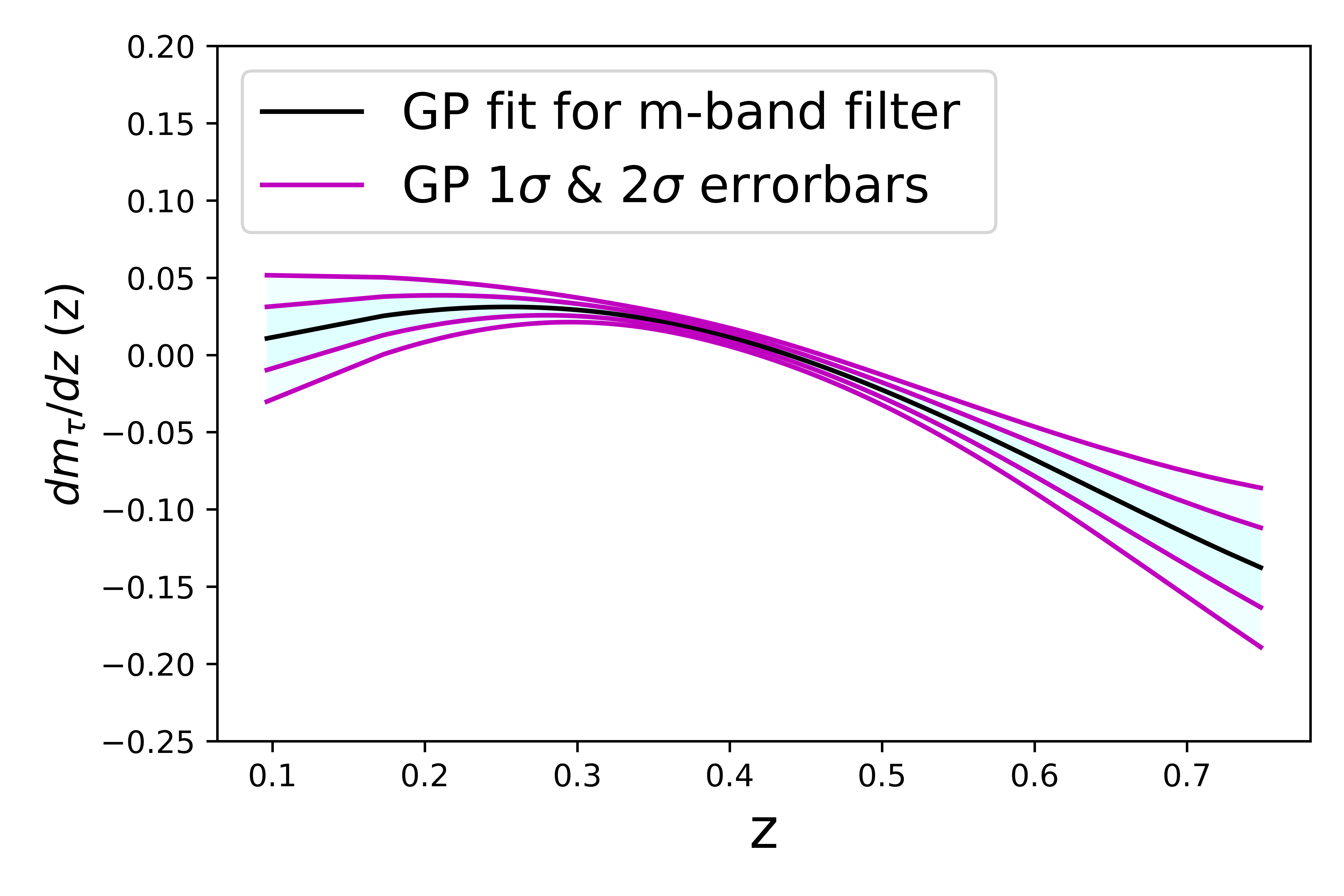}
  \caption{$\frac{d m_\tau}{dz}$ vs $z$ measurement for z filter}
  \label{fig:sub23}
\end{subfigure}

\caption{ The plot provides the measure of $\frac{d m_\tau}{dz}$ vs $z$  for all four filters estimated by the non-parametric smoothing technique,the Gaussian Process. The solid dark line indicates the best fit line and the colored lines across it highlights the $1\sigma$ and $2\sigma$ confidence region for the respective filter.}
\label{mdotdz}
\end{figure*}

This kernel is known as the square exponential kernel and as the simplest and default kernel for GP which has two hyper-parameters ($\sigma_f$ \& $l$). The output variance parameter, $\sigma_f$, determines the average distance of the function from its mean while the lengths scale, $l$, fixes the lengths of the wiggles in the smoothing function. The best fit value of these hyper-parameters is obtained by maximizing the corresponding marginal log-likelihood probability function of the distribution. In this paper, we utilized this method to fit the $m_\tau$ vs $z$ data points and to further find out the derivatives. The reconstructed plots of $\dfrac{d m_\tau}{dz}$ vs $z$  for all four filters using the Gaussian Process are shown in Fig \ref{mdotdz}.

\subsection{Obtaining $H(z)$ from Cosmic Chronometers}

The Hubble parameter $H(z)$ is one of the most crucial cosmological parameters which gives a measure of the many dynamical properties of the universe such as the evolution history, expansion, and composition of the universe. It is a pivotal parameter that even helps in understanding the nature of dark energy. The most precise and model independent method of estimating $H(z)$ at different redshifts is based on the differential ages of passively evolving galaxies. The Hubble parameter $H(z)$ can be expressed in terms of the rate of change of cosmic time with the redshift, given by
	
	\begin{equation}
	H(z)= - \dfrac{1}{(1+z)} \dfrac{\Delta z}{\Delta t}.
	\end{equation} 
	
\begin{figure*}[]
\centering
\includegraphics[width=0.7\linewidth]{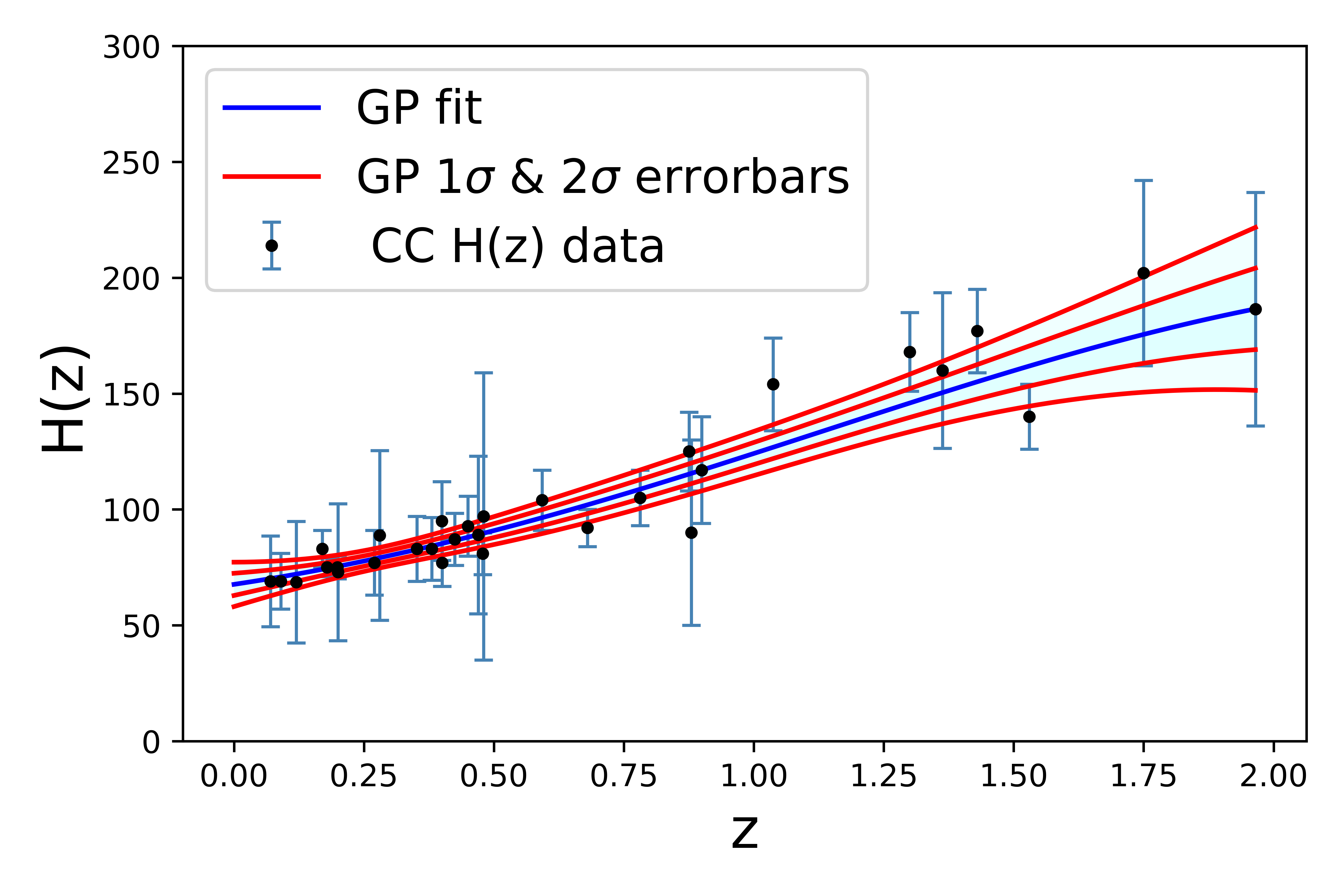}
\caption{ This plot contains the 31 datapoints of CC H(z) vs redshift with associated errorbars. Solid blue line indicated the best fit line and red lines represents the $1\sigma$ and $2\sigma$ error bars estimated using Gaussian Process.}
\label{fig:Part_II_P4}
\end{figure*}

	The change of cosmic time with redshift can be estimated from the ageing of the stellar population in galaxies. This method of calculating $H(z)$ is usually known as the ``Cosmic Chronometers'' and data points are generally referred to as CC $H(z)$. The most recent data compilation of the Hubble parameter measurements based on cosmic chronometers  has $31$ measurements of  $H(z)$ \citep{2018MNRAS.476.1036M, Wang_2017}. Again to reconstruct the value of the Hubble parameter $H(z)$ at all redshifts of the supernovae used in this analysis, we used the non-parametric method Gaussian Process. The corresponding plot can be seen in Fig. \ref{fig:Part_II_P4}.
	
\section{\label{section4}Result}

\begin{figure*}[t]
 \centering
 \begin{subfigure}{.5\textwidth}
   \centering
   \includegraphics[width=1.0\linewidth]{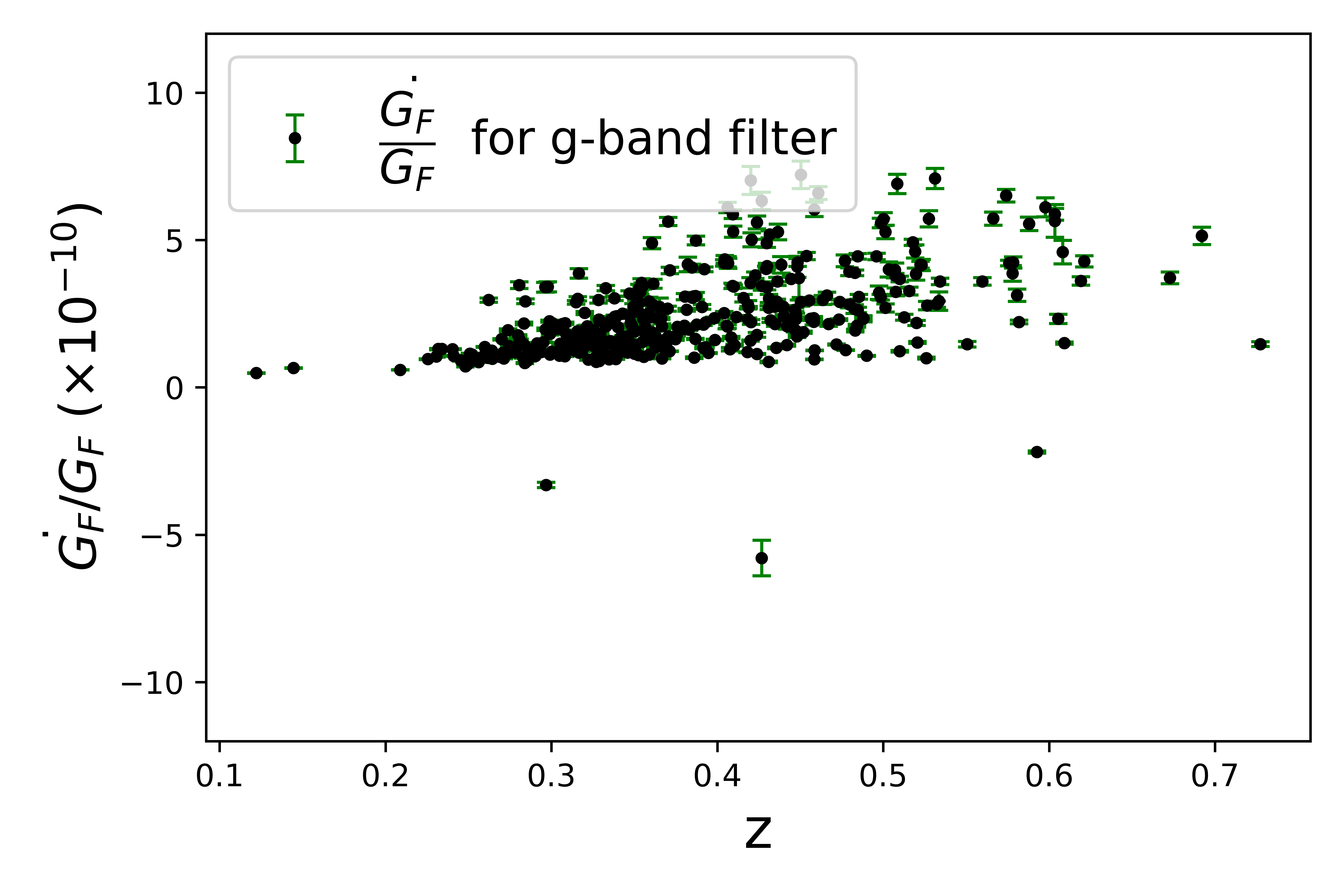}
   \caption{ $\dfrac{\dot{G}_F}{G_F}$ vs $z$ measurement for g filter}
   \label{fig:sub1z}
 \end{subfigure}%
 \begin{subfigure}{.5\textwidth}
   \centering
   \includegraphics[width=1.0\linewidth]{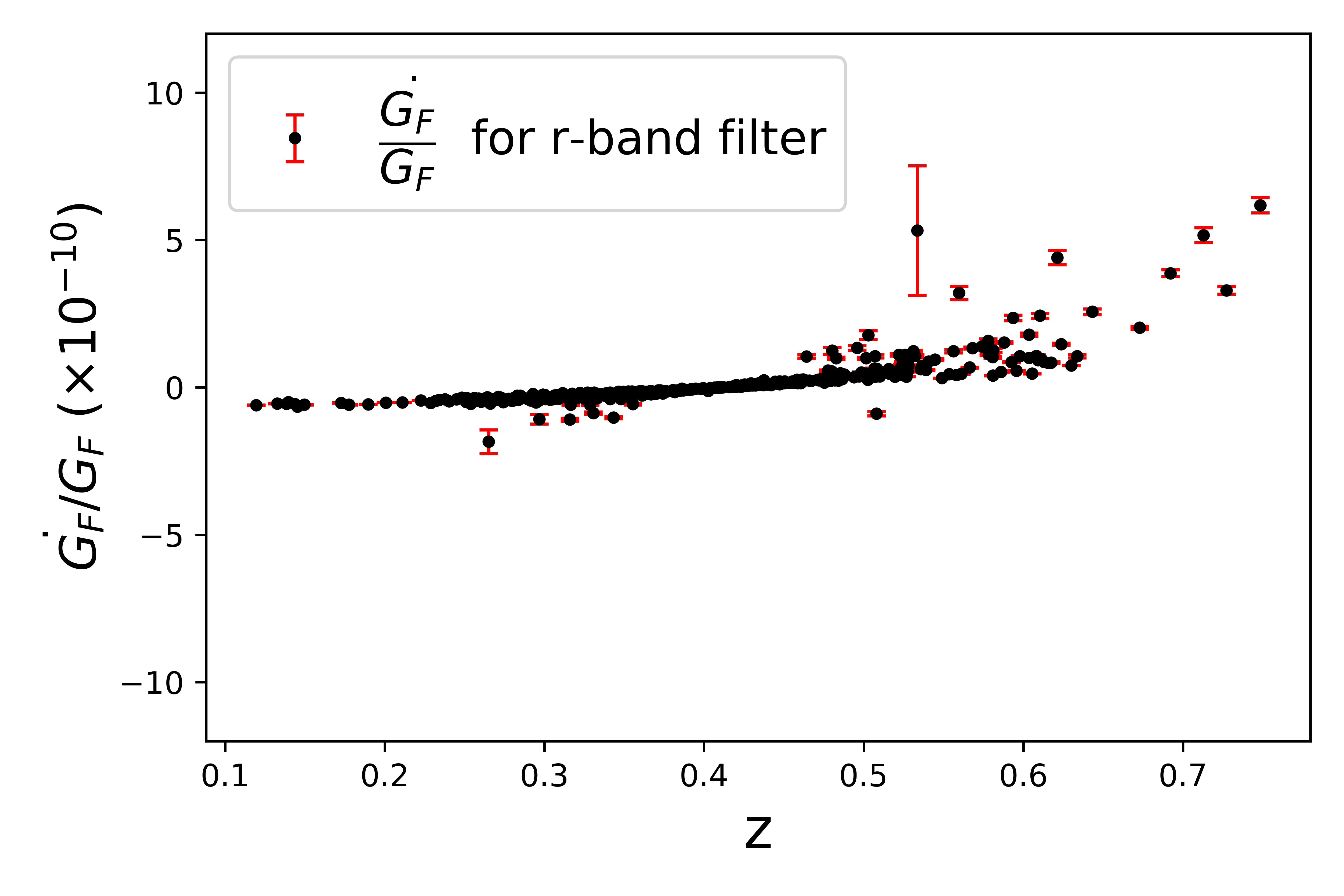}
   \caption{$\dfrac{\dot{G}_F}{G_F}$ vs $z$ measurement for r filter}
   \label{fig:sub24}
 \end{subfigure}

 \begin{subfigure}{.5\textwidth}
   \centering
   \includegraphics[width=1.0\linewidth]{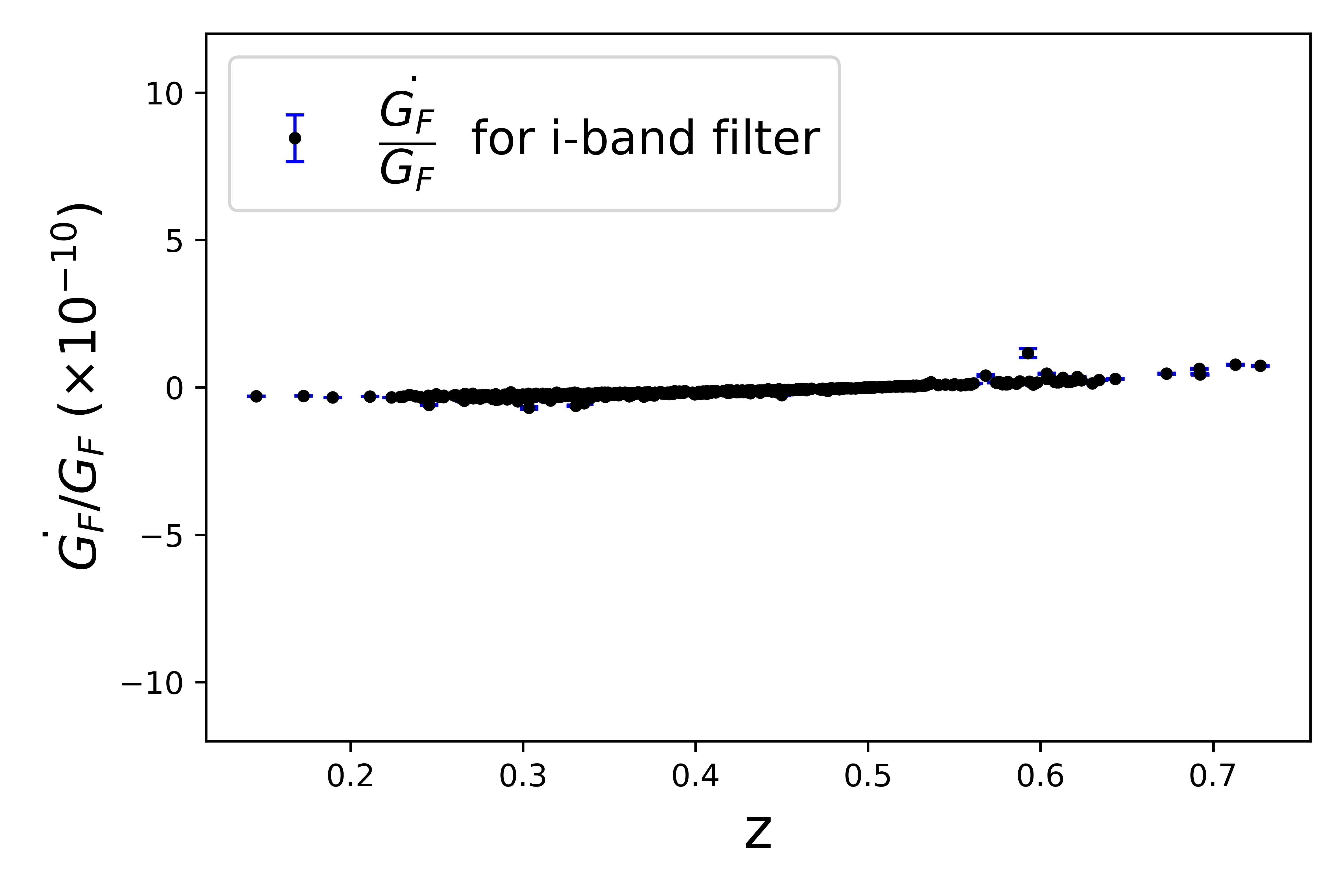}
   \caption{$\dfrac{\dot{G}_F}{G_F}$ vs $z$ measurement for i filter}
   \label{fig:sub13}
 \end{subfigure}%
 \begin{subfigure}{.5\textwidth}
   \centering
   \includegraphics[width=1.0\linewidth]{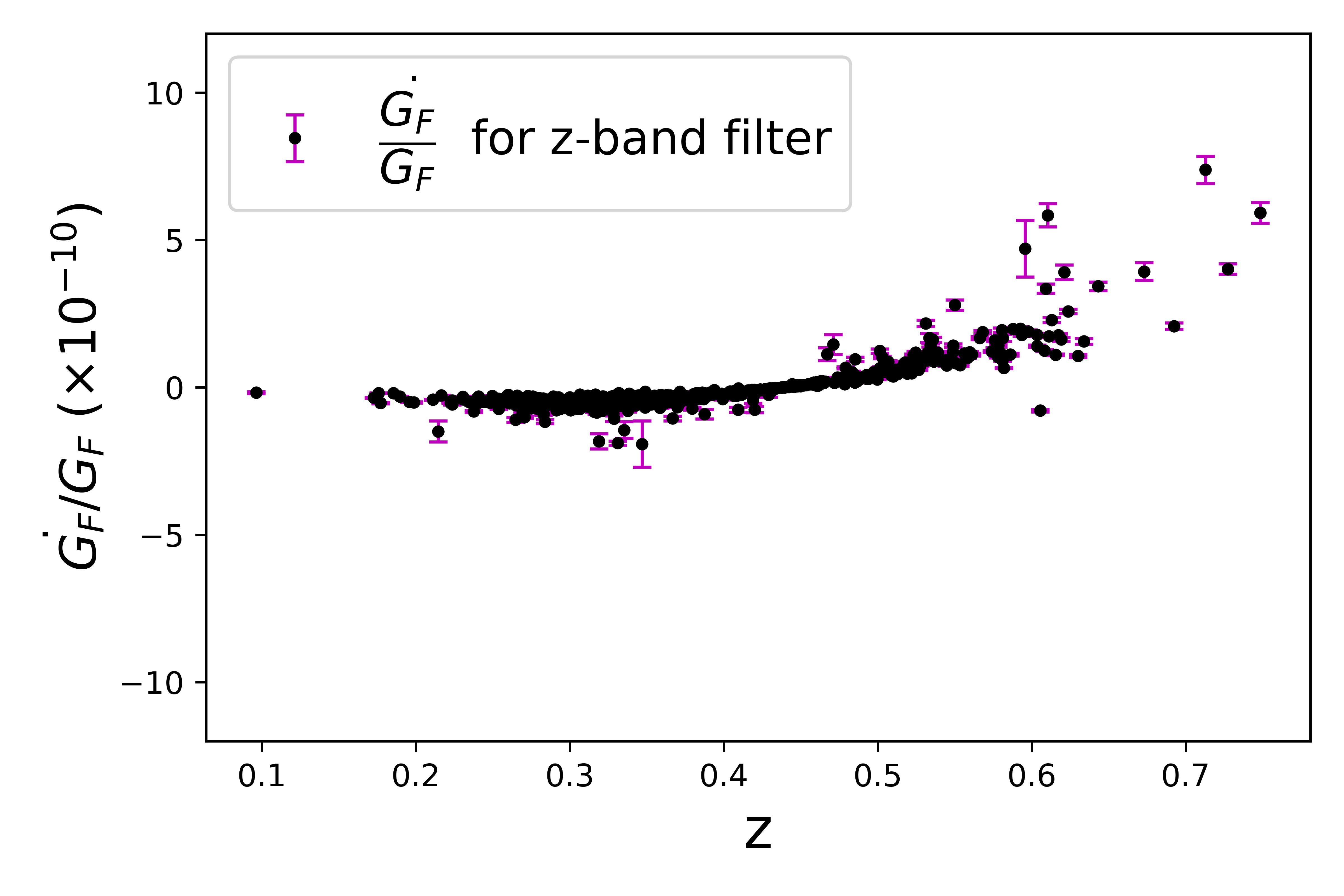}
   \caption{$\dfrac{\dot{G}_F}{G_F}$ vs $z$ measurement for z filter}
   \label{fig:sub25}
 \end{subfigure}

 \caption{These plots represent the reconstructed values of the fractional change  in the value of $G_F$ using light curve measurements of SNe Ia in four different filters.  }
 \label{gdotdz}
 \end{figure*}

In section \ref{section3}, we determined the value of $m_\tau$ and $\dfrac{d m_\tau}{dz}$ at different redshifts $z$ by using the SNe Ia light curve observation  in four different spectral filters. Besides that, the Hubble parameter $H(z)$ values are estimated using the cosmic chronometers, and reconstructed by using the non-parametric method, the Gaussian Process. Now, one can use the Eq.\ref{dm/dz} to estimate $\dfrac{\dot \Gamma}{\Gamma}$ and further Eq.\ref{maineq} to find out the fractional change in the Fermi coupling constant, $\dfrac{\dot G_F}{G_F}$. The estimated value of $\dfrac{\dot G_F}{G_F}$ corresponding to selected SNe Ia for four filters is shown in Fig. \ref{gdotdz}. The visual inspection of plots gives us a qualitative analysis yet it is important to do a proper quantitative analysis of the temporal variation of $\dfrac{\dot G_F}{G_F}$ with $z$.

\subsection{Temporal variation of $\dfrac{\dot{G}_F}{G_F}$}
\begin{figure*}
 \centering
 \begin{subfigure}{.5\textwidth}
   \centering
   \includegraphics[width=1.0\linewidth]{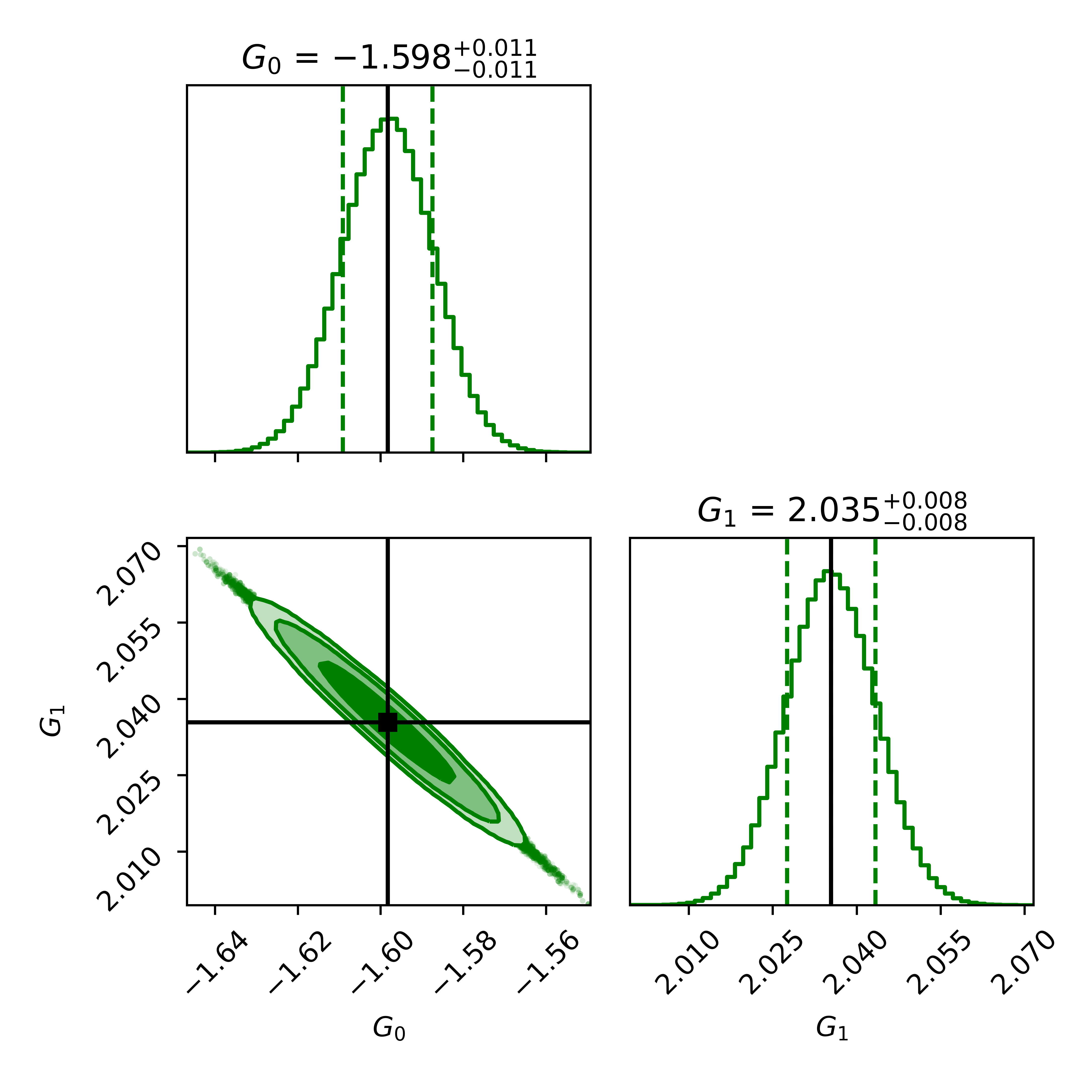}
   \caption{  g filter}
   \label{fig:sub1y}
 \end{subfigure}%
 \begin{subfigure}{.5\textwidth}
   \centering
   \includegraphics[width=1.0\linewidth]{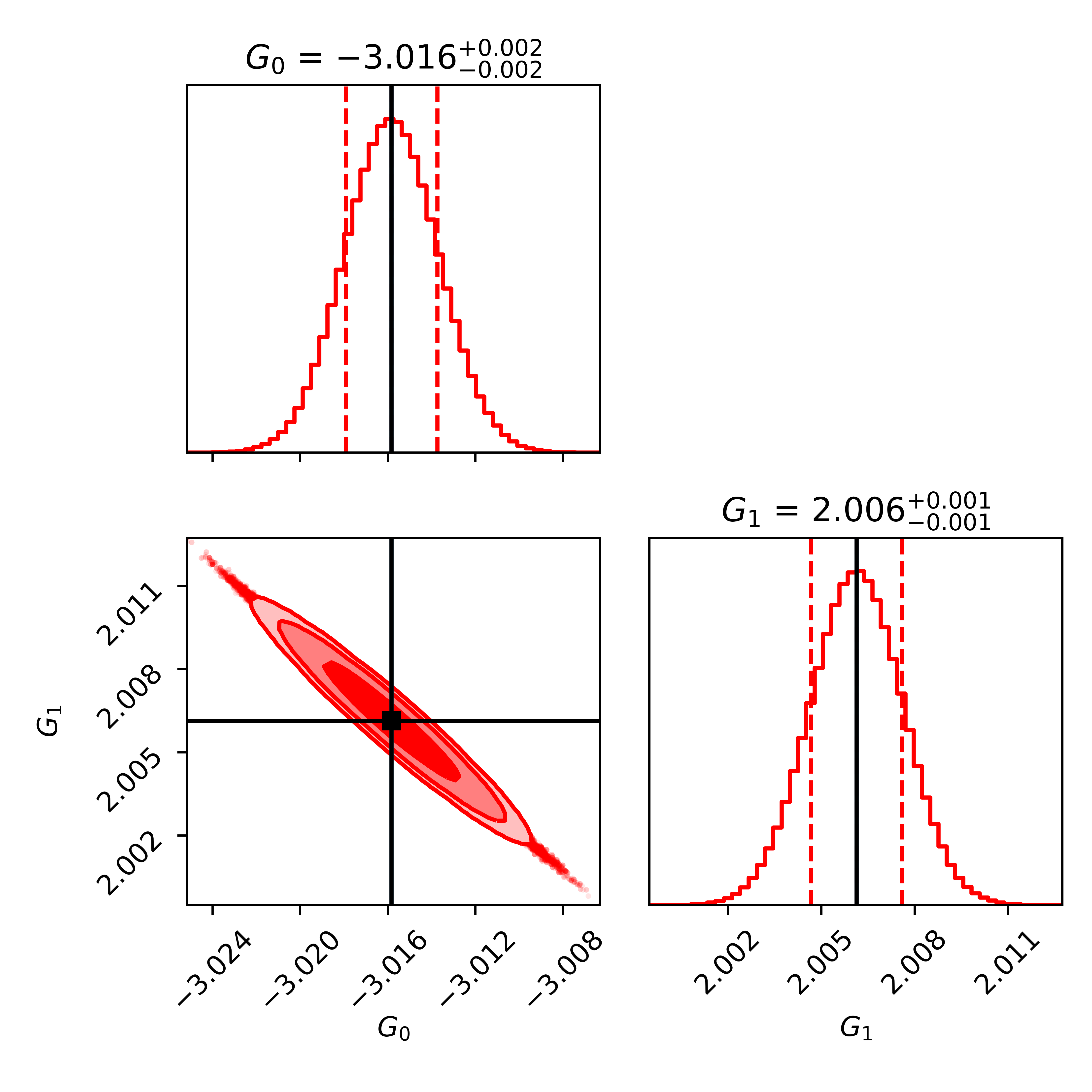}
   \caption{ r filter}
   \label{fig:sub26}
 \end{subfigure}

 \begin{subfigure}{.5\textwidth}
   \centering
   \includegraphics[width=1.0\linewidth]{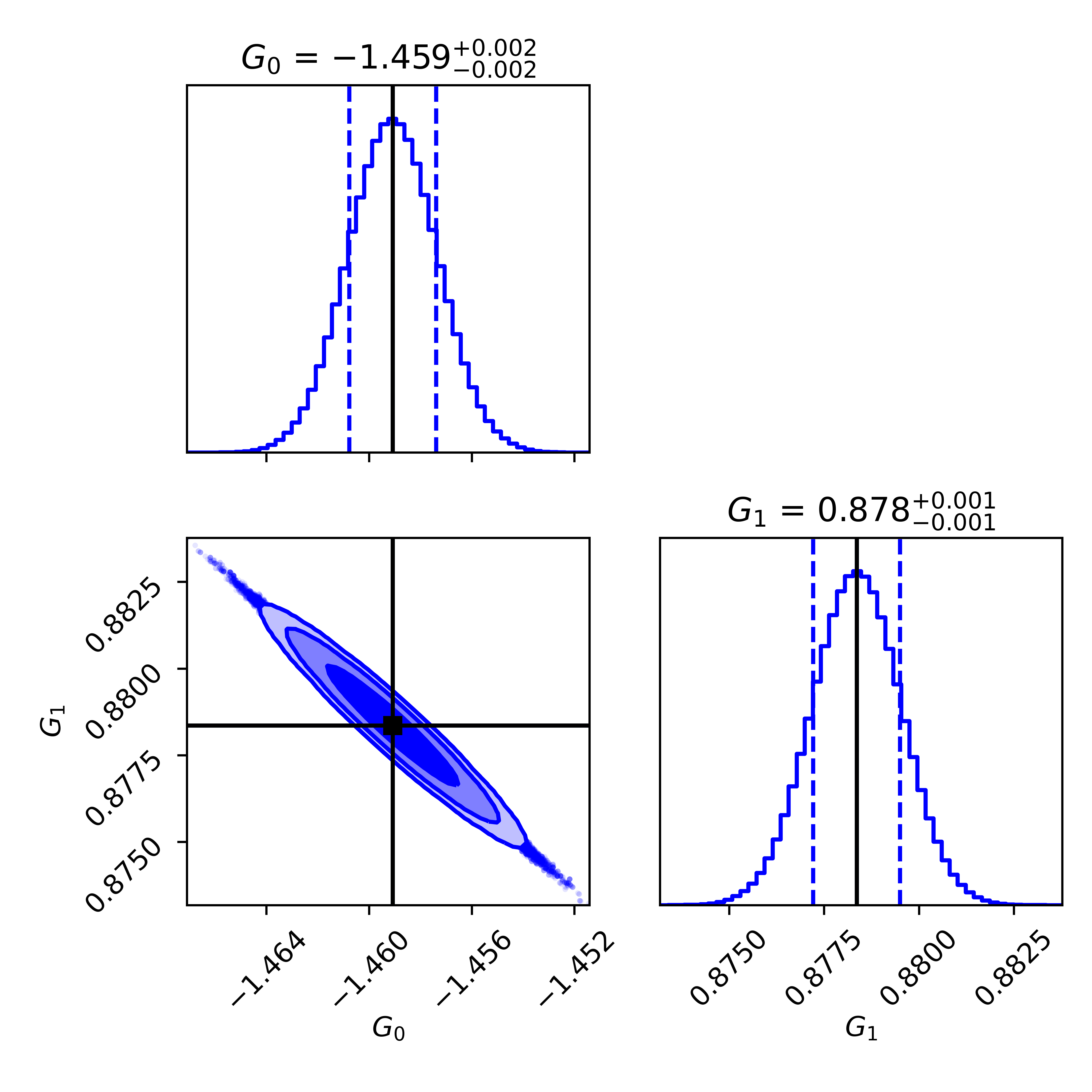}
   \caption{ i filter}
   \label{fig:sub14}
 \end{subfigure}%
 \begin{subfigure}{.5\textwidth}
   \centering
   \includegraphics[width=1.0\linewidth]{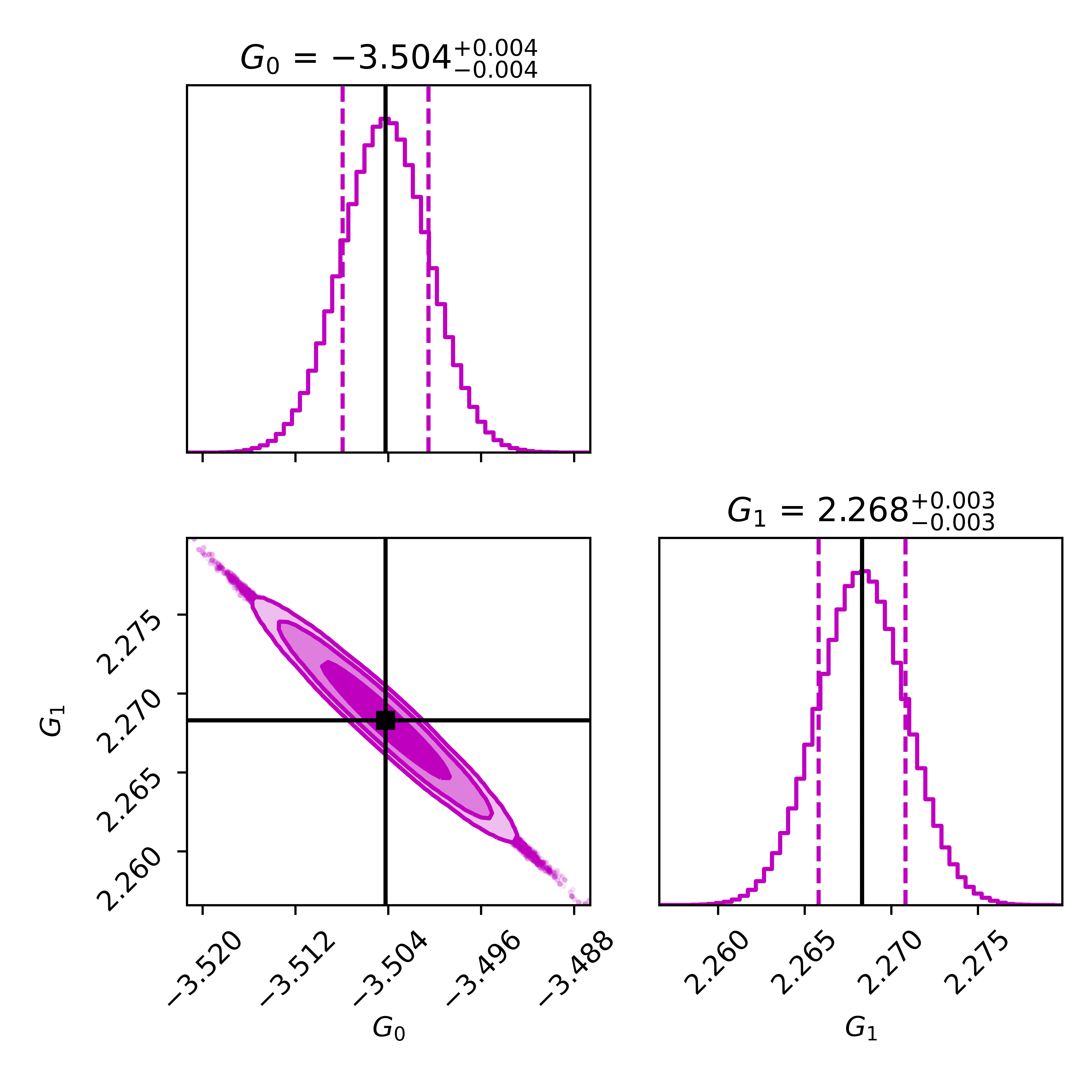}
   \caption{ z filter}
   \label{fig:sub27}
 \end{subfigure}

 \caption{ Standard emcee corner plot showing the one- and two-dimensional posterior distributions for model parameters $\dfrac{\dot{G}_F}{G_F} = G_0+G_1 e^z$ for all four filters.}
 \label{mdotdzfit}
\end{figure*}

In order to analyse any evolution of $\dfrac{\dot{G}_F}{G_F}$ with respect to $z$, we chose to fit a shifted exponentially increasing curve to the data derived in Fig.\ref{gdotdz}. The function is given as $\dfrac{\dot{G}_F}{G_F}(z) = G_0+G_1 e^z$ and is fitted using the Markov Chain Monte Carlo (MCMC) optimization techniques. For this, we used emcee package\cite{emcee}. The best fit values of the parameters and corresponding contours for different filters are shown in Table.\ref{tb:Part_II_P4} and Fig.\ref{mdotdzfit} respectively. While reading the result a constant factor of $10^{-10}$ must be taken care of. The function is chosen so that non-linear variation can also be studied without increasing the number of parameters(which would be needed for polynomial model functions). Further $\{1, e^z\}$ are chosen as a basis functions instead of $\{1, e^{-z}\}$ or $\{e^z, e^{-z}\}$ due to the visually clear convex increasing nature of the data in Fig. \ref{gdotdz}.

\begin{table*}[t]
 	\centering
 	\renewcommand{\arraystretch}{2}
 	\begin{tabular}[b]{| c | c |c|c|c|}\hline
 		Filter & $G_0 (\times 10^{-10})$ & $G_1 (\times 10^{-10})$ & $\frac{\dot{G}_F}{G_F} \big\rvert_{z=0}(\times 10^{-10})$ &$\frac{\dot{G}_F}{G_F} \big\rvert_{z=0.75}(\times 10^{-10})$ \\ \hline \hline
		
 		g-filter  & $-1.598^{+0.011}_{-0.011}$   & $2.035^{+0.008}_{-0.008}$ & $0.437^{+0.014}_{-0.014}$ & $2.710^{+0.020}_{-0.020}$\\ \hline
 		r-filter  & $-3.016^{+0.002}_{-0.002}$   & $2.006^{+0.001}_{-0.001}$ & $-1.010^{+0.002}_{-0.002}$ & $1.230^{+0.003}_{-0.003}$\\ \hline
 		i-filter  & $-1.459^{+0.002}_{-0.002}$    & $0.878^{+0.001}_{-0.001}$ & $-0.581^{+0.002}_{-0.002}$ & $0.399^{+0.003}_{-0.003}$\\ \hline
 		z-filter  & $-3.504^{+0.004}_{-0.004}$ & $2.268^{+0.003}_{-0.002}$& $-1.236^{+0.005}_{-0.004}$ & $1.297^{+0.007}_{-0.007}$\\ \hline
 	\end{tabular}
 	\caption{ This table contains the best fit values of $G_0$ and $G_1$ with $1\sigma$ confidence level. Along with this last two columns presents the value of $\frac{\dot{G}_F}{G_F}$ [$yr^{-1}$] at present $(z=0)$ and redshift $z=0.75$, respectively. }
 	\label{tb:Part_II_P4}
 \end{table*}

According to the results obtained in Table.\ref{tb:Part_II_P4}, the present value of fractional change in the Fermi coupling constant i.e. $\dfrac{\dot{G}_F}{G_F} \big\rvert_{z=0} = G_0 + G_1$ suggests that for the r and z-filter filters $G_0+G_1$ is of the order of $\approx 10^{-10}$.  However, for the g and i-filter filter analysis, $G_0+G_1$ values emerge to be of the order of $\approx 10^{-11}$ putting a very strong constraint on the present value of fractional change in $G_F$. An inconsistency is noticed in the g-filter which gives a positive value as opposed to the other filters which are all negative. For all filters we consistently find that the Fermi coupling constant increases with redshift as its fractional change varies exponentially. As all four filters correspond  to the same set of Type Ia supernovae light curves hence we expect them to be consistent and the  inconsistency in the g-filter might be an artifact of the variation in the error associated with the measurements using different filters. This error is reflected in the anomalous measurements of $m_{\tau}$ and subsequently $\dfrac{d m_{\tau}}{dz}$ for g-filter as well. Finally, our analysis suggests that for the i-filter this analysis puts the strongest upper bound on the present value of fractional change in the Fermi coupling constant value i.e.
  
\begin{equation}
     \dfrac{\dot{G}_F}{G_F} \bigg|_{z=0} \approx \left(-5.81^{+0.003}_{-0.003} \right) \times 10^{-11} \text{yr}^{-1} \hspace{1cm} \text{(for i-filter)}
 \end{equation}

This value has been estimated by using the observable distributed over a redshift range of $0<z<0.75$. Interestingly, all four filters suggests an increase in the value of $\dfrac{\dot{G}_F}{G_F}$ at the redshift $z=0.75$. Further, for $i$ filter, the value of $\dfrac{\dot{G}_F}{G_F} \big\rvert_{z=0.75}$ turns up to be of the order $10^{-11} \text{yr}^{-1}$. It is important to highlight that this is the upper bound on the fractional change in the value of $G_F$ over time and it doesn't rule out the possibility of this constant being a fundamental constant. Instead, improved, stringent, and stronger upper bounds further support the constancy of $G_F$. 

It is crucial to highlight that our analysis is based on several key assumptions which can be responsible for this small yet finite deviation of the present value of fractional change in $\dfrac{\dot{G}_F}{G_F}$  from zero. In the analysis, $G_F$ hasn't been measured directly instead it is derived from the independent measurements of SNe Ia light curves. Though, the studies suggest that the initial phases of the SNe Ia light curve are completely governed by the radioactive decay of $Ni^{56}$ and $Co^{56}$, any change in the inherent composition of different supernovae can potentially impact the overall light curve and decay rate. However, we are doing a statistical analysis over a large number of observables with very stringent selection criteria hence any such anomaly in the analysis would average out 

\section{\label{section5}Discussion}

In this work, we presented a novel approach to constrain the temporal variation of the Fermi coupling constant using the decay rates of primordial nucleosynthesis measured from the dimming of light curves of SNe Ia obtained from the PAN-STARRS supernovae catalog. Specifically, we measure the time variation of SNe Ia apparent magnitude up to $\approx 60$ days after the peak which we relate to the electroweak decay rate of $Ni^{56} \to Co^{56}$. Key points of our discussion are as follows:

\begin{itemize}

    \item We probe the Fermi coupling constant for a redshift range of $0<z<0.75$ for a maximum of 622 SNe Ia in four different spectral filters (\textit{griz}). Our analysis puts a stringent constraint on the present value of the fractional change in the Fermi coupling constant, i.e; $ \dfrac{\dot{G}_F}{G_F} \bigg|_{z=0} \approx 10^{-11} \text{yr}^{-1}$.
    
    \item  The analysis suggest the increase in the present value of $ \dfrac{\dot{G}_F}{G_F}$ for $r$, $i$ \& $z$ filters. However, $g$ filter suggests a negative value of this fractional change. We expect consistency in the results of the different filters for both $G_0$ and $G_1$, hence the mismatch in the calculations for the $g$ filter appear as an anomaly and can be attributed to the result of the error associated with the measurements in different filters. 
    
    \item Interestingly, at redshift $z=0.75$, all four filters consistently suggests the positive cosmic evolution of the Fermi coupling constant and the  strongest upper bound on the fractional change further strengthen to be $10^{-11} \text{yr}^{-1}$ for $i$ filter. 
    
    \item  Independent of filters, it is crucial to discuss the other primary sources of error that may have impacted our analysis. The first one could be the assumption of uniformity in the nature of the radioactive decay across all SNe Ia itself.  However, given the present models and simulations of SNe Ia explosion, dynamics and evolution it appears to be a reliable assumption to hold \cite{Hoflich:2003bg, ch1,ch2, ni,Diehl_2014}. 
    
    \item Further, to find out the peak of the SNe Ia light curves we used the result provided by the SNANA algorithm and re-verified our values using the SNCOSMO algorithm (both based on the SALT2 algorithm of SNe Ia light curve fitting which takes care of relative K-corrections while analysing)\cite{SNCOSMO, SNANA, guy2007, Guy2010Nov, NO2}. As the peak of the light curve is hidden in a sea of noisy points incorrect isolation of the light curve peak can lead to outlier points and large error bars.  Besides it, the large error bars in data may also contribute significant error in regression parameters and further may impact the determination of $\dfrac{\dot G_F}{G_F}$. 
    
    \item An unlikely error can be introduced if SNe Ia light curves are not identified with significant confidence . To avoid this source of error, we applied a very strong selection criterion on the well established \textit{PSNID classifier} for selecting only the light curves corresponding to SNe Ia.  Further, to refrain from having any significant statistical error and to avoid any cosmological model dependence in our analysis, we used a very precise and widely used non-parametric smoothing technique namely, Gaussian Process, to reconstruct the Hubble parameter value and also to find out the first derivative of $m_{\tau}$ with redshift \cite{rasmussen_gp,seikel_gp,gp2}.

\end{itemize}

 While prone to these systematic and associated random errors, our analysis manages to provide the most stringent upper bound on the present value of the  temporal variation in $G_F$. Constraining the Fermi coupling constant further implies constraints on the vacuum expectation value of the Higgs boson and hence on the masses of fundamental particles including specifically the mass of the $W$-boson, $M_W$. The strong constraint this study places on $G_F$ further confirms the constancy of fundamental particle masses hence agreeing with the Standard Model. 

 The analysis can further be improved by considering a larger dataset of SNe Ia with more precise measurement at higher redshifts and by having better understanding of SNe Ia light curve physics. Although our analysis places strong bounds on $\dfrac{\dot G_F}{G_F}$, it only covers SNe up to $z<0.75$, hence extending the analysis for higher redshift would be interesting. Additionally, the scope for further research includes probing the spatial variation of $G_F$ using a similar approach.
 
 \begin{acknowledgments}
 The authors wish to thank the Principal, St. Stephen's College, and the Centre of Theoretical Physics, St. Stephen's College for providing a platform to pursue research. AR is thankful to Prof. Deepak Jain, Prof. Debajyoti Choudhury, and Dr. David Jones for providing useful suggestions which lead to significant improvement to the draft. AR is also thankful to Prof. Shobhit Mahajan and ICARD,University of Delhi for providing research facility in the Department of Physics and Astrophysics, University of Delhi. 
 \end{acknowledgments}

\nocite{*}

\bibliography{apssamp}

\end{document}